\documentclass[10pt, twocolumn, twoside]{IEEEtran}

\usepackage{cite} 
\ifCLASSINFOpdf 
    \usepackage[pdftex]{graphicx}
\else
    \usepackage[dvips]{graphicx}
\fi
\usepackage{amsmath} 
\usepackage{algorithmic} 
\usepackage{array} 
\usepackage{url} 
\usepackage{epsfig}
\usepackage{multirow}
\usepackage[usenames, dvipsnames]{color}
\usepackage{subfig}
\usepackage{enumerate}
\usepackage{CJKutf8}
\usepackage{amsfonts}
\usepackage{comment}
\usepackage{algorithm}
\usepackage{hyperref}
\usepackage{diagbox}
\usepackage{slashbox}
\usepackage{mathtools, nccmath}
\DeclarePairedDelimiter{\nint}\lfloor\rceil
\usepackage{commath}
\usepackage[table,xcdraw]{xcolor}
\usepackage{textcomp}
\usepackage{float}

\usepackage{pgfplots}
\usepackage{tikz}
\usepackage{paralist}
\usepackage{xcolor}
\usetikzlibrary{
    patterns,
}
\usepgfplotslibrary{
    groupplots,
}

\pgfplotsset{compat=1.17}
\begin{document}

\title{TERA: Self-Supervised Learning of \\ Transformer Encoder Representation for Speech}

\author{Andy~T.~Liu, 
        Shang-Wen~Li, 
        and~Hung-yi~Lee 
\thanks{\copyright~2021 IEEE. Personal use of this material is permitted. Permission from IEEE must be obtained for all other uses, in any current or future media, including reprinting/republishing this material for advertising or promotional purposes, creating new collective works, for resale or redistribution to servers or lists, or reuse of any copyrighted component of this work in other works.}
\thanks{Accepted Version. Manuscript received July 12, 2020; revised February 25, 2021 and June 1, 2021; accepted June 29, 2021. Date of publication July 8, 2021; date of current version July 30, 2021. This work was supported in part by the National Science Council, Taiwan, under Contract 110-2628-E-002-001. The work of A. Liu was supported in part by the Frontier Speech Technology Scholarship of National Taiwan University and in part by ASUS AICS. The associate editor coordinating the review of this manuscript and approving it for publication was Dr. Ozlem Kalinli. (Corresponding author: Hung-yi Lee.).}
\thanks{Andy T. Liu and Hung-yi Lee are with the Graduate Institute of Communication Engineering, College of Electrical Engineering and Computer Science, National Taiwan University, Taipei 10617, Taiwan (e-mail: \href{mailto:f07942089@ntu.edu.tw}{f07942089@ntu.edu.tw}; \href{mailto:hungyilee@ntu.edu.tw}{hungyilee@ntu.edu.tw}).} 
\thanks{Shang-Wen Li is with Amazon AI, New York, NY 10001 USA (e-mail: \href{mailto:swdanielli@gmail.com}{swdanielli@gmail.com}).} 
\thanks{A. Liu was supported by Frontier Speech Technology Scholarship of National Taiwan University. A. Liu was also supported by ASUS AICS.} 
\thanks{Digital Object Identifier: \url{http://dx.doi.org/10.1109/TASLP.2021.3095662}}
}




\maketitle

\begin{abstract}
We introduce a self-supervised speech pre-training method called TERA, which stands for Transformer Encoder Representations from Alteration. Recent approaches often learn by using a single auxiliary task like contrastive prediction, autoregressive prediction, or masked reconstruction. Unlike previous methods, we use alteration along three orthogonal axes to pre-train Transformer Encoders on a large amount of unlabeled speech. The model learns through the reconstruction of acoustic frames from their altered counterpart, where we use a stochastic policy to alter along various dimensions: time, frequency, and magnitude. TERA can be used for speech representations extraction or fine-tuning with downstream models. We evaluate TERA on several downstream tasks, including phoneme classification, keyword spotting, speaker recognition, and speech recognition. We present a large-scale comparison of various self-supervised models. TERA achieves strong performance in the comparison by improving upon surface features and outperforming previous models. In our experiments, we study the effect of applying different alteration techniques, pre-training on more data, and pre-training on various features. We analyze different model sizes and find that smaller models are strong representation learners than larger models, while larger models are more effective for downstream fine-tuning than smaller models. Furthermore, we show the proposed method is transferable to downstream datasets not used in pre-training.
\end{abstract}


\IEEEpeerreviewmaketitle

\section{Introduction}
\label{sec:intro}

\IEEEPARstart{U}{nlike} humans, capable of self-learning through experiences and interactions, current real-world speech applications like automatic speech recognition (ASR) rely heavily on large amounts of human annotations.
For the next generation of speech processing systems to exhibit similar cognitive intelligence levels as humans, machines should be designed to learn from unlabeled data as humans do.
In the era of big data, self-supervised learning has emerged as an attractive approach to leverage knowledge from many unlabeled data, and are shown effective for improving downstream systems~\cite{cpc, wav2vec, vq_wav2vec, vq_wav2vec_ft, wav2vec2,  bidir_cpc, modified_cpc, apc1, apc2, improved_apc, vq_apc, decoar, audio2vec_1, audio2vec_2, wavenet_autoencoder, vc, phase_predict, pase, convDMM, mockingjay, audioalbert, speech_encoder, speechxlnet, mpc, mpc2, mpe, npc}. 

In self-supervised learning, an auxiliary task (or pre-training task) is formulated, and models are trained to solve it.
While solving the auxiliary task, the network is learning a function that maps input to desired representations.
Hence the fundamental tenet of self-supervised learning is the design of an auxiliary task, which allows the model to leverage knowledge from unlabeled data.
As such, the formulation of the auxiliary task should be carefully chosen.
The task should be challenging enough for the model to learn high-level semantic properties and not be too amiable to exploit low-level shortcuts.

After self-supervised pre-training, learned models could be applied to downstream Speech and Language Processing (SLP) tasks through feature-based \textit{speech representation} extraction, or \textit{fine-tuning} as part of the downstream model.
Speech representations are compact vectors which aim to capture high-level semantic information from raw speech~\cite{cpc, wav2vec, vq_wav2vec, bidir_cpc, modified_cpc, apc1, apc2, improved_apc, vq_apc, decoar, audio2vec_1, audio2vec_2, wavenet_autoencoder, vc, phase_predict, pase, convDMM, mockingjay, audioalbert, npc}. 
Thus, the goal of \textit{speech representation} learning is to find a transform that maps the input acoustic features into such vectors.
When the pre-trained networks are re-used as features, it provides a useful speech representation to reduce classifier complexity, makes high-level information more accessible, and ultimately improves downstream SLP tasks.
Besides, speech representations also help transfer learning and adaptation across different data distributions~\cite{bidir_cpc, modified_cpc, apc2, mockingjay, audioalbert}.
On the other hand, the \textit{fine-tuning} approach uses the pre-trained model to initialize a downstream model for supervised training.
The parameters of self-supervised learned models are found to be good initialization for ASR~\cite{vq_wav2vec_ft, wav2vec2, mockingjay, audioalbert, speech_encoder, speechxlnet, mpc, mpc2, mpe}.

\begin{figure}[t]
\centering
    \includegraphics[width=\linewidth]{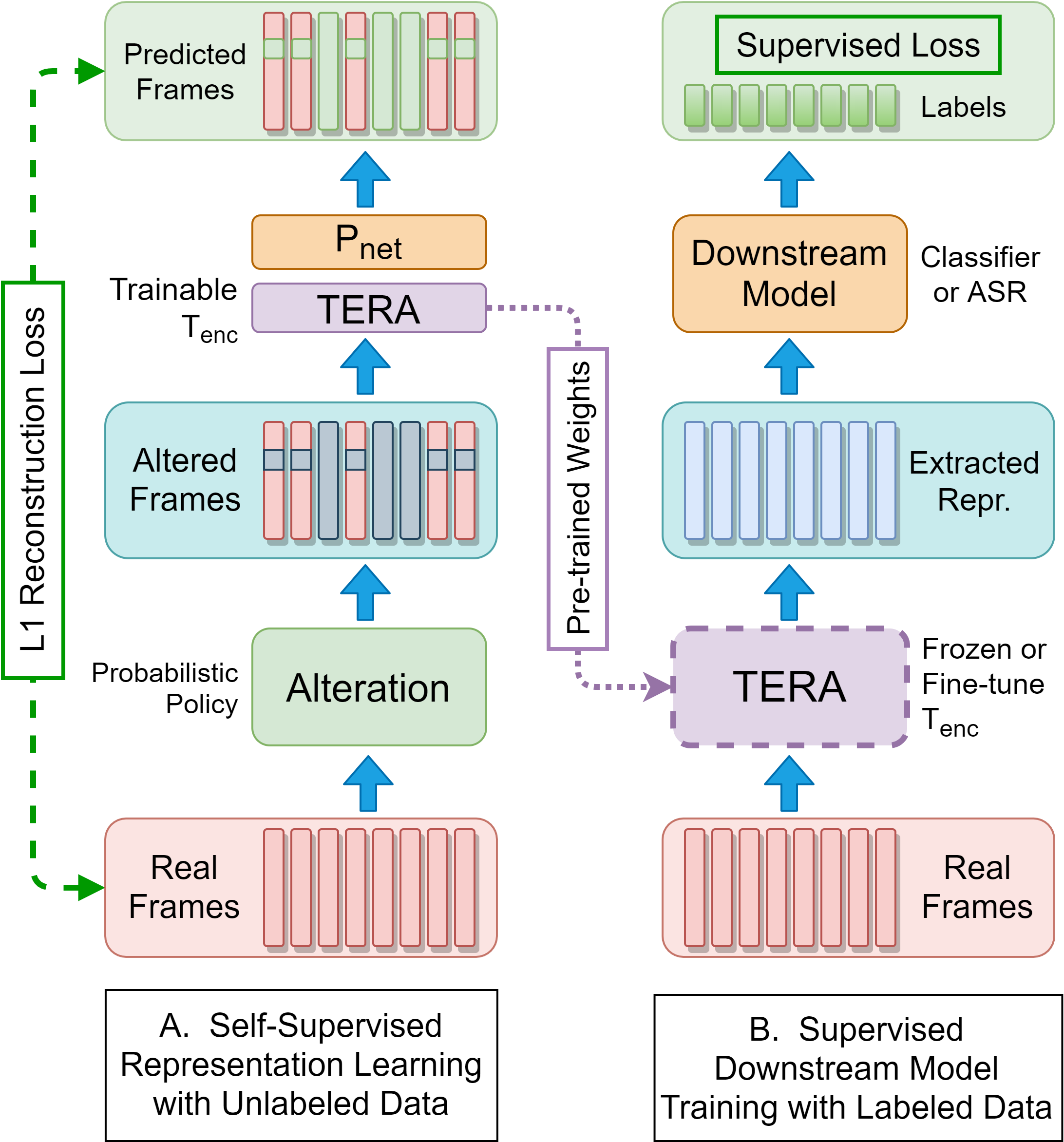}
    \caption[font=small]{The illustration of the proposed TERA self-supervised speech representation approach. \vspace{-5mm} }
\label{fig:proposed}
\end{figure}

In this work, we propose TERA: Transformer Encoder Representations from Alteration, where we use alteration on data to pre-train Transformer Encoders~\cite{transformer}.
We introduce a total of three types of alteration to form the self-supervised pre-training scheme: 
1) time alteration: reconstructing from corrupted blocks of time steps.
2) frequency alteration: reconstructing from missing blocks of frequency bins.
3) magnitude alteration: reconstructing from altered feature magnitudes.
These alterations can be applied together or separately in the pre-training process.
We apply alteration on data by dynamically sampling through a probabilistic policy to create random alterations.
The model acquires information about the content around the corrupted or altered portions, and by reconstructing them, the model learns a more contextualized representation.
We illustrated the framework in Fig.~\ref{fig:proposed}.

We use the following downstream tasks to evaluate TERA: phoneme classification, keyword spotting, speaker recognition, and automatic speech recognition (ASR).
Also, we compare the effectiveness of each alteration method separately and in combination.
As a result, we confirm that each of the proposed alteration methods guides the model to learn a distinct aspect of speech:
1) The time alteration effectively enforces a more accurate phoneme prediction, keyword detection, and speech recognition, as it leads the model to learn richer phonetic content.
2) The frequency alteration effectively improves speaker prediction accuracy, as it leads the model to encode speaker identity.
3) The magnitude alteration effectively improves performance for all tasks, as it potentially increases data diversity for pre-training.

Different self-supervised frameworks have been widely studied, in Section~\ref{sec:related} we provide a thorough review.
Previous work explored mostly for reconstruction on the temporal axis, for example unidirectional (or autoregressive) reconstruction of magnitude or phase from past frames \cite{apc1, apc2, improved_apc, wavenet_autoencoder, vc, phase_predict, pase, speechxlnet}, or bidirectional reconstruction of a temporal frame from both past and future slices \cite{mockingjay, audioalbert, decoar, audio2vec_1, audio2vec_2, speech_encoder, mpc, mpc2, mpe, npc}.
Our work contrasts with prior work in several ways.
Firstly, unlike previous work that only employs reconstruction on the temporal axis, we use reconstruction loss and apply alteration on data along three orthogonal axes, including time, frequency, and magnitude axis.
Secondly, most works evaluated their approach with classification tasks only~\cite{cpc, modified_cpc, apc1, vq_apc, wavenet_autoencoder, vc, phase_predict, audio2vec_1, audio2vec_2, mockingjay, audioalbert}.
In contrast, we moved beyond classification and applied our model to ASR.
For a comprehensive investigation, we evaluate our method with four downstream tasks.
Thirdly, we explore knowledge transfer between pre-trained models and downstream tasks, an under-investigated problem in speech compared to NLP~\cite{nlp_transfer2, nlp_transfer3}. 
We leverage two ways to incorporate the pre-trained model with downstream tasks, where most of the previous work only explored one way of transferring their pre-trained models.
Fourthly, we study how self-supervised models behave when pre-trained on a different amount of unlabeled data.
Surprisingly, we find that methods that learn from time-only masked reconstruction methods can sometimes not benefit from more unlabeled data.
This is because of the reconstruction nature, memorizing all the details, including the unnecessary noise.
Additionally, we study the effect of pre-training on various features.
We explore four different acoustic features in this work, including log Mel, fMLLR, MFCC, and FBANK.
We report results primarily on log Mel and fMLLR.
The usage of fMLLR, is not explored before for reconstruction-based methods.
Also, none of the previous work explores more than one acoustic feature for their method.
Our study finds that using different acoustic features in reconstruction-based learning has a significant effect on pre-trained models and is a parameter choice for researchers.
Furthermore, we find smaller pre-trained models perform well for feature extraction over larger models, and larger models tend to be more effective for fine-tuning.
Finally, we show explicitly that our approach continues to work well in the face of domain mismatch between pre-training and downstream datasets.
For reproducibility of our results, we provide our implementation with pre-trained models and evaluation scripts in the S3PRL Toolkit\footnote{\href{https://github.com/s3prl/s3prl}{\texttt{The S3PRL Toolkit: https://github.com/s3prl/s3prl}} \label{github}}~\cite{S3PRL}.

\begin{figure*}[ht]
\centering
    \includegraphics[width=\textwidth]{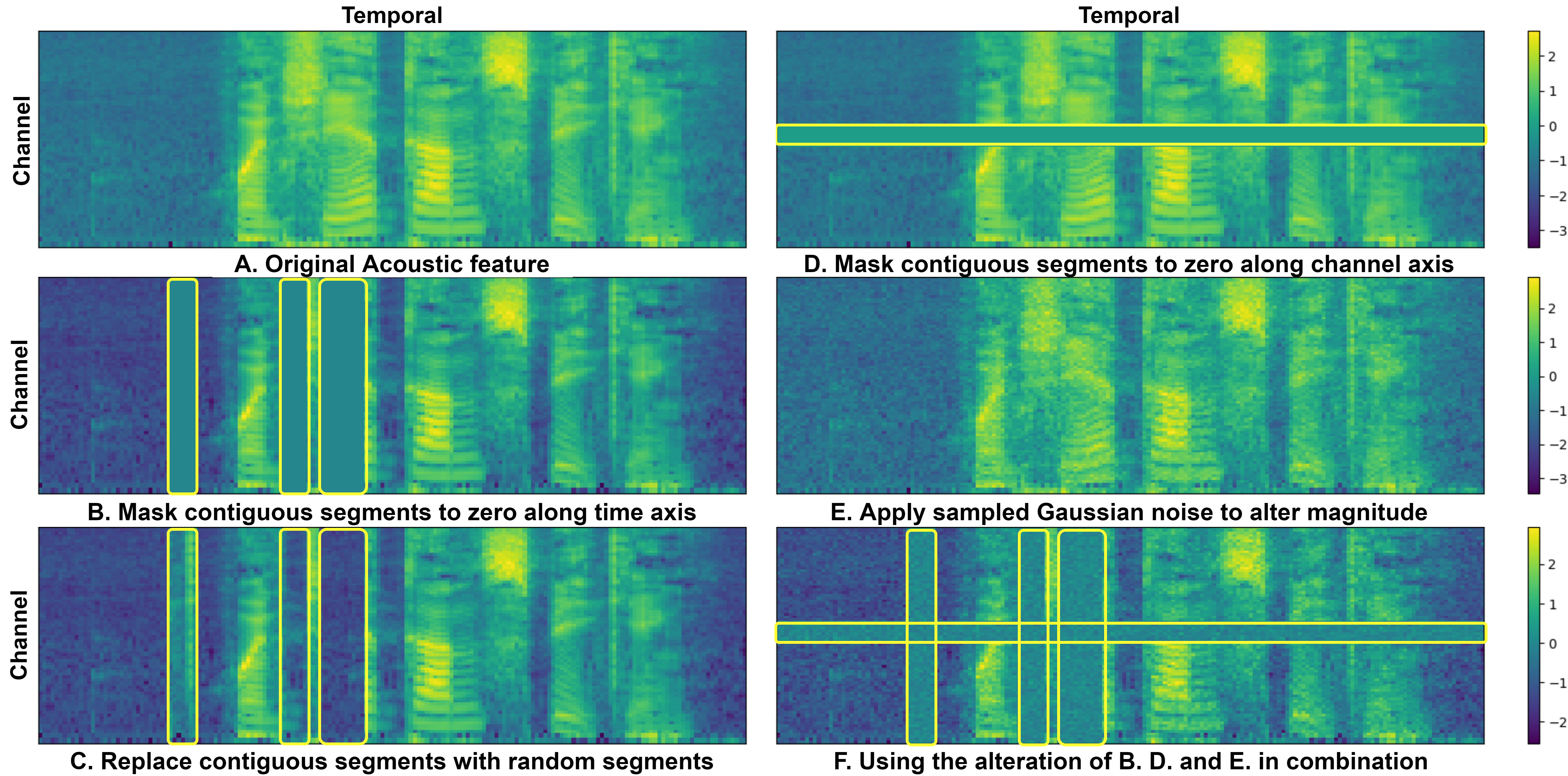}
    \caption[font=small]{The illustration of different inputs with various alteration applied for the proposed auxiliary objective. The altered part is highlighted in \textcolor{yellow}{yellow}. \vspace{-5mm} }
\label{fig:input}
\end{figure*}

\section{Related Work}
\label{sec:related}
There are two major branches of speech pre-training methods: Contrastive Predictive Coding (CPC) and Reconstruction.

\subsection{Contrastive Losses}
\label{ssec:cpc}

\subsubsection{Contrastive Predictive Coding}
\label{sssec:cpc}
The CPC paper~\cite{cpc} describes a form of unidirectional modeling in the feature space, where the model learns to predict the near future frames in an acoustic sequence while contrasting with frames from other sequences or frames from a more distant time.
In wav2vec~\cite{wav2vec}, the CPC~\cite{cpc} loss is used to pre-train speech representations for speech recognition, and experiment results show self-supervised pre-training improves supervised speech recognition.

\subsubsection{CPC with Quantization}
\label{sssec:wav2vec}
In the work of vq-wav2vec~\cite{vq_wav2vec}, the wav2vec~\cite{wav2vec} approach is incorporated with the well-performing Natural Language Processing (NLP) algorithm -- Bidirectional Encoder Representations from Transformers (BERT)~\cite{bert, roberta}.
The vq-wav2vec~\cite{vq_wav2vec} approach learns BERT-like speech representations through a two-stage training pipeline.
In a follow-up work of BERT + vq-wav2vec~\cite{vq_wav2vec_ft}, the pre-trained vq-wav2vec~\cite{vq_wav2vec} model is directly fine-tuned on transcribed speech using a Connectionist Temporal Classification (CTC)~\cite{ctc} loss instead of feeding the representations into a task-specific model.
In wav2vec 2.0~\cite{wav2vec2}, the vq-wav2vec method is improved to a single-stage training scheme through time masking in the latent space.



\subsubsection{Improving CPC}
\label{sssec:improving_cpc}
In recent research, the CPC loss has also been extended and applied to bidirectional context networks.
Bidirectional CPC (Bidir-CPC)~\cite{bidir_cpc} representations are learned from bidirectional predictive models.
The multi-layer CNN encoder network is shared for both directions, but two autoregressive GRU context networks read encoded observations from the forward and backward contexts.
The Modified CPC~\cite{modified_cpc} focuses on improving the CPC approach by introducing several modifications to the original.
These modifications include changing the batch normalization to channel-wise normalization, replacing the linear prediction layer with a Transformer layer~\cite{transformer}, and replacing the GRU context network with LSTM cells.

In this work, TERA is compared with several contrastive learning methods, including CPC~\cite{cpc}, wav2vec~\cite{wav2vec}, vq-wav2vec~\cite{vq_wav2vec}, BERT + vq-wav2vec~\cite{vq_wav2vec_ft}, wav2vec 2.0~\cite{wav2vec2}, Bidir-CPC~\cite{bidir_cpc}, and Modified CPC~\cite{modified_cpc}.

\subsection{Reconstruction Losses}
\label{ssec:reconstruction}
Another recently emerged branch of speech pre-training approach devotes its attention on reconstruction losses.

\subsubsection{Autoregressive reconstruction}
\label{sssec:autoregressive}
Primarily inspired by language models (LM) for text, the Autoregressive Predictive Coding (APC)~\cite{apc1, apc2} model can be seen as a speech version of LM.
The APC approach uses an autoregressive model to encode temporal information of past acoustic sequence; the model then predicts future frames like a recurrent-based LM~\cite{rnn} while conditioning on past frames.
In \cite{improved_apc}, the APC objective is extended to multi-target training.
The new objective predicts not only the future frame conditioning on previous context but also past memories.
In VQ-APC~\cite{vq_apc}, a vector quantization (VQ) layer is used with the APC objective, which imposes a bottleneck and forces the model to learn better representations.
In DeCoAR~\cite{decoar}, combining the bidirectionality of ELMo~\cite{elmo} and the reconstruction objective of APC~\cite{apc1, apc2}, models were able to learn deep contextualized acoustic representations.

\subsubsection{Time-only Masked Reconstruction}
\label{sssec:time-only-masked}
Largely inspired by the Masked Language Model (MLM) task from BERT~\cite{bert, roberta, albert} and Permutation Language Modeling (PLM) from XLNet~\cite{xlnet}, recent work~\cite{mockingjay, audioalbert, speech_encoder, speechxlnet, mpc, mpc2, mpe, npc} have explored using BERT-style tasks to pre-train speech encoders.
These approaches adopt the NLP pre-training technique to continuous speech.
In Mockingjay~\cite{mockingjay}, input frames of speech are masked to zero to pre-train Transformer Encoders.
In Audio ALBERT~\cite{audioalbert}, Mockingjay is modified to have shared parameters across Transformer layers.
In \cite{mockingjay_defense}, Mockingjay is shown to be effective in defending adversarial black-box attacks.
And in \cite{understanding}, the self-attention of Mockingjay is shown to be meaningful and explainable.
On the other hand, TERA can be seen as an improved version of Mockingjay~\cite{mockingjay}.
Using the time alteration alone as pre-training objective reduces TERA to Mockingjay.

In \cite{mpc, mpe}, time-only masked reconstruction following the standard BERT masking policy is employed to pre-train ASR encoders.
In \cite{mpc2}, a simpler masking policy is employed, where input features are divided into chunks of four frames, and masking on chunks are applied with a probability of 15\%.
In Speech-XLNet~\cite{xlnet}, models learn by reconstructing from shuffled input speech frame orders rather than masked frames.
In \cite{speech_encoder}, SpecAugment~\cite{spec_augment} is applied on input frames to pre-train ASR encoders (bi-GRUs).
In wav2vec 2.0~\cite{wav2vec2}, time masking is applied in the latent space.
In Non-autoregressive Predictive Coding (NPC)~\cite{npc}, time masking is introduced through Masked Convolution Blocks, rather than on the input data.
In \cite{slu_bert}, phoneme posterior vectors are used to train a standard BERT~\cite{bert, xlnet} model.
The phoneme posterior vectors are output from a supervised acoustic model, which requires CTC loss training over the ground-truth phonemes.
Also, in \cite{bertphone}, CTC loss is used along with time-only masked reconstruction training to learn phonetic representations.
As \cite{slu_bert} and \cite{bertphone} both use phoneme labels for CTC training, they diverge from other works that are fully self-supervised.

\subsubsection{Learning from Other Reconstruction Losses}
\label{sssec:autoencoder}
Other than autoregressive and time-only masked reconstruction losses, previous works have also explored the reconstruction of different targets or frameworks, including temporal slice estimation, gap estimation, autoencoders, phase prediction, and Markov Models.
In Audio2Vec~\cite{audio2vec_1, audio2vec_2}, the model learns through reconstructing a spectrogram slice from past and future slices; this can be seen as a speech version of the NLP Word2Vec~\cite{word2vec} variants CBoW (continuous bag-of-words) and skip-gram.
The TemporalGap~\cite{audio2vec_1, audio2vec_2} approach learns through estimating the temporal gap between two short audio segments extracted at random from the same audio clip.
In \cite{wavenet_autoencoder}, speech representations are learned by applying autoencoding neural networks to speech waveform.
Apart from reconstructing spectrograms, in \cite{phase_predict}, representations are learned through reconstructing the phase of the short-time Fourier transform from its magnitude.
In PASE~\cite{pase}, a single neural encoder learns to solve multiple self-supervised tasks at once, including reconstruction of waveform, Log power spectrum, MFCC, prosody, and other binary discrimination tasks.
The ConvDMM~\cite{convDMM} approach learns speech representations with convolutional neural networks and Markov Models.
Although The design of the auxiliary task fundamentally decides what the model learns through its reconstruction.

In this work, TERA is compared with several reconstruction learning methods, including APC~\cite{apc1, apc2}, VQ-APC~\cite{vq_apc}, DeCoAR~\cite{decoar}, Mockingjay~\cite{mockingjay}, Audio ALBERT~\cite{audioalbert}, SpecAugment~\cite{speech_encoder, spec_augment}, and NPC~\cite{npc}.


\section{Proposed Methodology}
\label{sec:proposed_method}

\subsection{Alteration on Data}
\label{ssec:alterations}

As illustrated in Fig.~\ref{fig:proposed}A, the input acoustic frames (outlined in the red box) and target predicted frames (outlined in the green box) could be any acoustic features, such as MFCC, FBANK, fMLLR, or log Mel.
We show a sample of 80-dimensional log Mel feature sequence from the LibriSpeech~\cite{librispeech} \textit{train-clean-100} subset in Fig.~\ref{fig:input}A.
We denote the entire speech corpus as $\mathcal{X}$ and the acoustic features of the utterance sampled from $\mathcal{X}$ as $\overrightarrow{x}$.
The length (the number of frames) and the height (the number of frequency bins) of $\overrightarrow{x}$ is denoted as $L_x$ and $H_x$, respectively.
Below, we introduce how we use different methods to alter the input $\overrightarrow{x}$.

\subsubsection{Time Alteration}
\label{sssec:time}
Our model learns bidirectional representations from past and future contexts by altering contiguous segments along the time axis.
In time alteration, a certain percentage of input frames are altered during training, and the model attempts to reconstruct the corrupted span from neighboring frames.
We randomly select $T_{num}$ amount of starting indexes $I_T$ without replacement to alter the input utterance.
The amount $T_{num}$ is given as the maximum time alteration percentage $P_T$ normalized by the time alteration width $W_T$:
\begin{equation}
    T_{num} = \nint{ P_T \times L_x \div W_T}
    \label{eq:num_index}
\end{equation}
Note that if time alteration width $W_T = 1$, then $T_{num} = P_T \times L_x$.
For each starting index location $i_t$ in $I_T$, we alter $W_T$ consecutive frames from $i_t$ according to the following stochastic alteration policy: 
1) 80\% of the time, we mask all the selected frames to zero.
2) 10\% of the time, we replace all with random segments of frames.
3) For the rest 10\% of the time, we do nothing and leave the frames in $\overrightarrow{x}$ unchanged.
The design of Case 3) is to allow the model to receive real inputs during training and addresses the train-test inconsistency problem.
This inconsistency problem comes from the absence of alteration during inference time, and the model will only receive acoustic features without alteration.

We illustrate the masking and replacing of frames in Fig.~\ref{fig:input}B and \ref{fig:input}C, respectively.
Our time alteration policy is more sophisticated than other time-only masked reconstruction approaches~\cite{speech_encoder, mpc2}, where they simply mask a percentage with zeroed-out spans, unlike ours that have random and real frames.
We set the time alteration width $W_T$ to 7 frames, which corresponds to 85ms of speech. The time alteration width $W_T$ then lies in the average phoneme duration range (average phone duration is around 50 ms to 100 ms at usual rates of 10 to 20 phones per second).
We set the $P_T$ percentage of total altered frames to 15\%, as suggested in \cite{mockingjay, bert, roberta}.
We allow time alteration blocks to overlap each other, hence resulting in the larger highlighted yellow box in the left of Fig.~\ref{fig:input}B and \ref{fig:input}C.
With overlapping, we generate a longer altered span ($> W_T$) and force the model to infer on more global structure rather than a fixed local span ($W_T$).
The idea behind time alteration is that a model that can predict the partial loss of small speech segments should provide a contextualized understanding of previous and future content.
Our experiments show that the proposed time alteration is the essential element that drives models to learn bidirectional understanding, resulting in a substantial improvement compared to models that are not using the time alteration method.

\subsubsection{Frequency Alteration}
\label{sssec:frequency}
Our second pre-training method is frequency alteration. 
It is largely inspired by SpecAugment proposed for ASR augmentation~\cite{spec_augment}, and the ASR pre-training scheme proposed in \cite{speech_encoder}. 
We randomly mask the values of a block of consecutive frequency bins to zero for all time steps across the input sequence for this alteration.
The block of masked frequency is selected by first sampling the width of block $W_C$ from $\{0, 1, . . . , W_C\}$ uniformly. 
Then, we sample a frequency index $I_C$ from $\{0, 1, . . . , H_x - W_c - 1\}$, where $H_x$ is the number of frequency bins in input sequence $\overrightarrow{x}$.
The frequency bins from $I_C$ to $I_C + W_c - 1$ are those to be masked. 
Note that the policy will mask none of the frequencies for $1/(W_C+1)$ of the time.
Thus, from time to time, the model will receive inputs with all of the frequency information.
By allowing the model to receive real inputs again addresses the inconsistency between training and inference time.

We illustrate this alteration's effect on input sequence in Fig. ~\ref{fig:input}D.
Unlike the time alteration case, where we sample many blocks for alteration as visualized in Fig.~\ref{fig:input}B and \ref{fig:input}C, we only sample a single block for frequency alteration in each utterance.
The reason is that acoustic sequences can be arbitrarily long and temporally smooth~\cite{long}, while there are only a limited and fixed number of frequencies $H_x$.
Hence, we select multiple blocks for time alteration, but only one block for the frequency axis.
Following the work of \cite{speech_encoder}, we set the maximum frequency alteration width $W_C$ to 16 frequency bins (20\% of the 80-dimension feature).
The intuition behind frequency alteration is that a model that can predict the partial loss of frequency information should learn a high-level understanding along the frequency axis.

As we will show in our experiments, we find that using frequency alteration provides a more linearly spreadable speaker representation and a stronger speaker recognizer.
Surprisingly, encoding speaker information through this objective does not compromise phoneme classification or ASR performance but instead increases performance. 
This makes TERA not only suitable for tasks that only require speaker information (e.g. speaker recognition) and tasks that only require phonetic information (e.g. speech recognition), but also beneficial for tasks that requires both speaker and phonetic information at the same time (e.g. voice conversion~\cite{vc}).

\subsubsection{Magnitude Alteration}
\label{sssec:magnitude}
We introduce the third method, magnitude alteration, by applying sampled Gaussian noise to augment input sequences' magnitude with a probability $P_N$. 
For $P_N$ of the time, we sample a random magnitude matrix $\overrightarrow{z}$ of dimensions $L_x$ and $H_x$, which has the same shape as $\overrightarrow{x}$.
Each element in $\overrightarrow{z}$ is sampled from the normal distribution $\mathcal{N}$ with zero mean and 0.2 variance.
We then add $\overrightarrow{z}$ on top of the real frames of $\overrightarrow{x}$.
We show the effect of magnitude alteration in Fig.~\ref{fig:input}E, where we apply magnitude alteration to the original utterances $\overrightarrow{x}$.
By altering input magnitude, we potentially increase the amount of pre-training data (which is similar to the idea of data augmentation~\cite{spec_augment}).
Additionally, magnitude alteration offers another variation to all the `mask to zero` cases described in Section~\ref{sssec:time} (time alteration) and Section~\ref{sssec:frequency} (frequency alteration).
We illustrate this new `mask to noise` variation in Fig.~\ref{fig:input}F, where the selected blocks of time and frequency are now with random magnitudes instead of zeros.
Empirically, altering magnitude provides a performance benefit for all downstream tasks, thanks to the increased input data variety. Also, the gain from magnitude alteration is additive to that of other alteration techniques.

\begin{algorithm}[!htbp]
    \renewcommand{\algorithmicrequire}{\textbf{Input:}}
    \renewcommand{\algorithmicensure}{\textbf{Output:}}
    \caption{The TERA self-supervised pre-training algorithm}
    \label{alg:algorithm1}
    \begin{algorithmic}[1]
    \REQUIRE
        Dataset $\mathcal{X}$, total training steps $T_{steps}$, time alteration percentage $P_T$, time alteration width $W_T$, frequency alteration width $W_C$, magnitude alteration percentage $P_N$.
    \ENSURE
    Transformer Encoders parameters $\theta_{enc}$
    
    \STATE Initialize Transformer Encoders $T_{enc}$ parameters $\theta_{tenc}$\
    \STATE Initialize Prediction Network $P_{net}$ parameters $\theta_{pnet}$\
    \FOR{$t=1$ to $T_{steps}$}
        \STATE Sample speech utterance $\overrightarrow{x}$ from $\mathcal{X}$
        \IF{apply time alteration}
            \STATE {$L_x$ = number of frames in $\overrightarrow{x}$}
            \STATE {$T_{num}$ = $\nint{P_T \times L_x \div W_T}$}
            \STATE {$I_T$ = Sample $T_{num}$ amount of starting indexes}
            \STATE {Sample value $v$ from uniform distribution $\mathcal{U}(0, 1)$}
            \IF{$v < 0.8$}
                \FOR{all $i_t$ in $I_T$}
                    \STATE {$\hat{x} \leftarrow$ mask $W_T$ number of consecutive frames in $\overrightarrow{x}$ to zero}
                \ENDFOR
            \ELSIF{$v >= 0.8$ and $v < 0.9$}
                \FOR{all $i_t$ in $I_T$}
                    \STATE {$\hat{x} \leftarrow$ replace $W_T$ number of consecutive frames in $\overrightarrow{x}$ with randomly sampled consecutive frames from the same utterance}
                \ENDFOR
            \ELSE
                \STATE{Do nothing, $\hat{x} \leftarrow \overrightarrow{x}$}
            \ENDIF
        \ENDIF
        \IF{apply frequency alteration}
            \IF{apply time alteration}
                \STATE{Apply on top of previous alteration, $\overrightarrow{x} \leftarrow \hat{x}$}
            \ENDIF
            \STATE {$W_c$ = Sample a frequency width from $\mathcal{U}(0, W_C)$}
            \STATE {$H_x$ = number of frequencies in $\overrightarrow{x}$}
            \STATE {$I_C$ = Sample a frequency index from $\mathcal{U}(0, H_x-W_c)$}
            \STATE {$\hat{x} \leftarrow$ starting from $I_C$, mask $W_c$ consecutive frequency bins in $\overrightarrow{x}$ to zero}
        \ENDIF
        \IF{apply magnitude alteration}
            \IF{apply time or frequency alteration}
                \STATE{Apply on top of previous alterations, $\overrightarrow{x} \leftarrow \hat{x}$}
            \ENDIF
            \STATE {Sample value $v$ from uniform distribution $\mathcal{U}(0, 1)$}
            \IF{$v < P_N$}
                \STATE {Sample magnitude vector $\overrightarrow{z}$ from $\mathcal{N}(0, 0.2)$}
                \STATE {Apply noise $\hat{x} \leftarrow \overrightarrow{x} + \overrightarrow{z}$ to alter magnitude}
            \ELSE
                \STATE{Do nothing, $\hat{x} \leftarrow \overrightarrow{x}$}
            \ENDIF
        \ENDIF
        \STATE {Compute L1 reconstruction loss to update:
        \begin{equation}
            L_{rec}(\theta_{tenc}, \theta_{pnet}) = \norm{\overrightarrow{x} - P_{net}(T_{enc}(\hat{x}))}_1
        \end{equation}
        }
    \ENDFOR
\end{algorithmic}
\end{algorithm}

\subsection{Pre-training TERA}
\label{ssec:pretraining}
We use the three alteration techniques to formulate the TERA self-supervised pre-training task, where the model is required to minimize the reconstruction error of acoustic features given altered frames as input.
The proposed three alteration methods can be used separately or used together as a mixture, as shown in Fig.~\ref{fig:input}.
The complete pre-training process of TERA is described in Algorithm~\ref{alg:algorithm1}, where we show how three alteration on data are deployed through the proposed stochastic policy.
The stochastic policy dynamically samples random pattern based on the applied alterations every time we feed an input sequence to the model.
We denote the altered input as $\hat{x}$.
In the case where multiple alterations are applied on the same input, $\overrightarrow{x}$ is used again to represent carried-on alterations.
After input alteration, we feed $\hat{x}$ into the Transformer Encoders $T_{enc}$ and the prediction network $P_{net}$.
The architecture of $P_{net}$ is consist of a 2-layer feed-forward network with a hidden size of 768.
For $T_{enc}$, we use Transformer Encoders with a hidden size of 768, number of self-attention heads as 12, dropout rate of 0.1, and the intermediate feed-forward layer hidden size as 3072.
We primarily report results on three model sizes: \textit{base} (Layer=3, Parameters=21.3M), \textit{medium} (Layer=6, Parameters=42.6M), and \textit{large} (Layer=12, Parameters=85.1M).
The Transformer Encoders $T_{enc}$ and the prediction network $P_{net}$ are connected to reconstruct $\overrightarrow{x}$ from $\hat{x}$.

L1 reconstruction loss is then computed between input $\overrightarrow{x}$ and network output from $P_{net}$ to update the network parameters $\theta_{tenc}$ and $\theta_{pnet}$.
We use gradient descent training with mini-batches of size $32$ to find model parameters that minimize the L1 loss under the self-supervised pre-training task.
We employ the AdamW optimizer~\cite{adamW} for updating model parameters, where the learning rate is warmed up over the first $7\%$ of total training steps $T_{steps}$ to a peak value of $2e^{-4}$ and then linearly decayed.
Our pre-training setup can be accommodated in a single 1080Ti GPU with 11GB of memory.
Computational efficiency allows interested parties to easily train our model with their data without massive computational resources.
The models are trained with fixed total training steps $T_{steps}$ (details in Section~\ref{ssec:dataset}).
After pre-training, the parameters $\theta_{tenc}$ of the Transformer Encoders $T_{enc}$ are retained for downstream tasks, while the prediction network $P_{net}$ is discarded, as illustrated in Figure~\ref{fig:proposed}.

In this work, our models' input is an 80-dimensional log Mel spectrogram if not specified otherwise.
We also explore pre-training with other acoustic features, including 39-dimensional MFCC, 80-dimensional FBANK, and 40-dimensional fMLLR.
We extract all of the features with a window of 25 ms and a stride of 10 ms. 
We apply per-utterance CMVN (cepstral mean and variance normalization) to the features.
We set the total pre-training steps $T_{steps}$ of TERA as 200k and 1M for 100 hours and 960 hours of pre-training data, respectively.

\subsection{Incorporating with Downstream Tasks}
\label{ssec:downstream}
We investigate different ways to incorporate the learned TERA model to downstream tasks.

\subsubsection{Representation Extraction}
We first use the weighted sum technique to investigate which layer is best for representation extraction. We use a learnable weighted sum to integrate hidden states from all layers of the self-supervised model when training with downstream tasks, similar to the ELMO~\cite{elmo} approach. We then analyze the learned weights and find that for TERA, APC~\cite{apc1, apc2}, and vq-wav2vec ~\cite{vq_wav2vec}, the weighted sum always favors the last layer the most (we investigate with phoneme and speaker classification tasks).
Hence in this work, we extract representations from TERA’s deepest layer, which is essentially the hidden states of the last Transformer Encoder layer.
The extracted representation is fed to downstream classifiers as input and replacing surface features.
Parameters of TERA are frozen when training downstream tasks in this approach.
In later experiments, we use this approach if not specified otherwise.

\subsubsection{Fine-tuning}
The second approach is to fine-tune the TERA model with downstream models.
Here the output of TERA is connected to a downstream model of any kind, as illustrated in Fig.~\ref{fig:proposed}.
We then update the pre-trained TERA together with random initialized downstream models.
We denote this approach as \textit{fine-tune} to distinguish from representation extraction in later experimental tables and figures.

\section{Experimental Setup}
\label{sec:experiment}
We use four downstream tasks for evaluation: phoneme classification, speaker recognition, keyword spotting, and speech recognition.
We use publicly known and available settings for our downstream tasks to link our works with previous ones and allow easier comparison.
For all experiments, we train with fixed random seeds for consistency.

\subsection{Datasets}
\label{ssec:dataset}
We use a total of three datasets.
We consider three subsets of LibriSpeech for the pre-training of TERA: the \textit{train-clean-100}, the \textit{train-clean-360}, and the \textit{train-other-500} subset.
The three subsets add up to a total of 960 hours of data.
We also use the LibriSpeech~\cite{librispeech} dataset for downstream tasks of phone classification, speaker recognition, and speech recognition.
The \textit{train-clean-100} subset is used in the phoneme classification and speaker recognition task.
For ASR, we use the \textit{train-clean-100} as labeled data, and the \textit{dev-clean} and \textit{test-clean} subset are used for evaluation.
We use the TIMIT~\cite{timit} dataset to evaluate the transferability of pre-trained models.
We do not use the TIMIT dataset for pre-training, but we use it for downstream phoneme classification and ASR tasks.
We consider two subsets of TIMIT: the training set and the complete test set.
As is usually done, we derive the development set and the core test set from the complete test set, according to the Kaldi~\cite{kaldi} TIMIT recipe and \cite{timit_phone}.
We report results by training downstream tasks on the training set and evaluating with the development and core test sets.
For keyword spotting, we use the Speech Commands~\cite{speech_commands} dataset.
We consider two subsets of Speech Commands: the training set and the test set.
We derive the development set from the validation list provided in Speech Commands~\cite{speech_commands}.
We report results by training keyword spotting models on the training set, and evaluating on the development set and test set.
The Speech Commands dataset is also not used for pre-training.

\subsection{Phoneme Classification Setup}
\label{ssec:phone_setup}

\subsubsection{Phoneme Classification on LibriSpeech}
We measure the frame-wise phoneme prediction performance with classifiers trained on top of representations.
Following previous work~\cite{cpc, modified_cpc}, we adapt the common setup using 41 possible phoneme classes and the \textit{train-clean-100} subset of LibriSpeech~\cite{librispeech}.
To make the comparison as fair as possible, we use aligned phoneme labels and train/test split provided in the CPC ~\cite{cpc} and Modified CPC\cite{modified_cpc} literature.
We derive 10\% of the data from the training split and use it as the development set.
Following the previous work~\cite{cpc}, we utilize linear classifiers to measure the linear separability of phonemes.
We denote these classifiers as \textit{linear}.
Additionally, following the work of Modified CPC~\cite{modified_cpc}, we also report results from performing linear classification with a concatenation of 8 windows, which matches the average length of a phoneme.
We denote this type of classifier setting as \textit{linear concat}.
As not all the information encoded is linearly accessible, in addition to measuring linear separability, we also use classifiers with one single hidden layer, following the same settings as in \cite{cpc}.
We denote such setting as \textit{1-hidden}.

\subsubsection{Phoneme Classification on TIMIT}
For TIMIT, we obtain a phoneme label for each frame from the manual phoneme transcriptions provided in TIMIT.
We measure accuracy after mapping the 48 phonemes to the smaller set of 39 phoneme classes, as is customary for TIMIT since \cite{hmm} (the same applies to ASR on TIMIT).
Following the classifier settings for LibriSpeech, we also use the \textit{linear}, \textit{linear concat}, and \textit{1-hidden} classifiers on TIMIT.
For both LibriSpeech and TIMIT, we report phoneme classification accuracy (\%) on the test set.
We investigate the domain shift issue with this setup, where models are pre-trained on LibriSpeech then used on TIMIT for the downstream task.

\subsection{Keyword Spotting Setup}
\label{ssec:keyword_spotting}
We evaluate representations on the keyword spotting task.
We follow the standard setup~\cite{speech_commands},
where the keyword spotting task is a 12-class classification problem.
The test set provides equal numbers of instances for each class; hence class accounts for approximately 8.3\%.
To probe representations, we did not use complex models for this task, as performance may saturate.
Instead, we use a two hidden layer feed-forward classifier, with mean pooling overtime applied just before the output layer.
We report keyword classification accuracy (\%) on the test set.
This setup allows us to study the domain shift issue, where models are pre-trained on LibriSpeech then used on Speech Commands for the detection task.

\subsection{Speaker Classification Setup}
\label{ssec:speaker_setup}
We evaluate representations on speaker prediction tasks.
Following the common experiment setting in \cite{cpc, mockingjay, audioalbert}, we adopt the LibriSpeech~\cite{librispeech} \textit{train-clean-100} subset, which consists of 251 speakers.
We use the same train/test split as provided in \cite{cpc}.
We evaluate the pre-trained models with two types of speaker classification tasks, frame-wise linear classification, and utterance-wise linear classification.
For frame-wise speaker classification, the classifier predicts the speaker for each input frame.
We denote this experiment setting as \textit{speaker frame}.
As for utterance-wise classification, we average the representation of each utterance over time.
Then the classifier predicts speaker identity conditioning on the averaged vector.
We denote this setting as \textit{speaker utter}.
For both settings, we employ a simple linear classification model.
In \cite{cpc, audioalbert}, they also investigate these two speaker classification settings.
In general, the \textit{speaker frame} classification task is more difficult than \textit{speaker utter}, however \textit{speaker utter} is a more common scenario for speaker classification.
Also, good results in \textit{speaker frame} always imply good results in \textit{speaker utter}, but not the other way around.
Hence we report both \textit{speaker frame} and \textit{speaker utter} for completeness.
For both tasks, we report speaker classification accuracy (\%) on the test set.

\subsection{Hybrid DNN/HMM ASR Setup}
\label{ssec:asr_setup}
We evaluate the performance of ASR models on top of representations.
We employ the Hybrid DNN/HMM ASR modeling implemented with the PyTorch-Kaldi~\cite{pytorchkaldi} toolkit.
We investigate two types of DNN settings, \textit{MLP} and \textit{liGRU}.
\textit{MLP} is a simple single-layer multilayer perceptron model.
\textit{liGRU} is a 5-layer light-gated recurrent unit followed by 2-layers of fully-connected network.
In first-pass decoding, we use a 4-gram language model with a beam-search algorithm.
The WER score is computed with the \textit{NIST SCTK} scoring toolkit~\cite{pytorchkaldi}.
For lattice rescoring, we use the Kaldi \textit{ConstArpaLm} format language model.
The decoding and rescoring scripts are inherited from the Kaldi \textit{s5} recipe~\cite{kaldi}.

When we utilize TERA as a representation extractor, we feed the output of TERA to \textit{liGRU} or \textit{MLP} and freeze the parameters of TERA during training.
As for fine-tuning TERA, we update the TERA model with \textit{liGRU} or \textit{MLP} as part of the DNN component in the hybrid ASR framework.
For our ASR experiments, we pre-train TERA on 40-dimensional fMLLR~\cite{fmllr} features instead of 80-dimensional log Mel features since the PyTorch-Kaldi ASR toolkit~\cite{pytorchkaldi} works best with fMLLR inputs.
To highlight the effect of pre-training, we use a limited amount of labeled data, i.e., the \textit{train-clean-100} subset, for supervised ASR training.
Hyperparameters are tuned on the \textit{dev-clean} subset, and testing results measured from the \textit{test-clean} subset are reported.
We report ASR modeling results of LibriSpeech in terms of WER.
In addition to evaluating with LibriSpeech, we also benchmark ASR results with TIMIT, where we measure PER.
With this setup, we investigate the domain shift issue, where models are pre-trained using data in the domains different than the downstream tasks.

\begin{figure*}[ht]
    \centering
    
    \begin{tikzpicture}
        \begin{axis}[
            ybar,
            ymin = 40,
            ymax = 93,
            y label style ={at={(axis description cs:0.50,1.1)}, anchor=center, rotate=-90},
            ylabel={Phoneme classification on seen data - LibriSpeech (41-classes)},
            extra y ticks = {65},
            extra y tick labels = { Accuracy (\%) },
            extra y tick style = { tick label style={rotate=90, yshift=7mm,},},
            bar width = 0.55cm,
            enlarge x limits = 0.25,
            height = 3.5cm,
            symbolic x coords = {linear, linear concat, 1-hidden},
            xtick=data,
            x=5.0cm,
            nodes near coords,
            nodes near coords align = {vertical},
            nodes near coords style = {
                font=\footnotesize,
                /pgf/number format/.cd,
                fixed,
                fixed zerofill,
                precision=1,},
            ]
            \addplot [pattern=crosshatch dots, pattern color=blue,]
            	coordinates {(linear, 45.81) (linear concat, 56.51) (1-hidden, 53.66)};
            \addplot [pattern=crosshatch dots, pattern color=red,]
            	coordinates {(linear, 63.50) (linear concat, 70.96) (1-hidden, 71.73)};
            \addplot [pattern=crosshatch dots, pattern color=orange,]
            	coordinates {(linear, 63.01) (linear concat, 70.57) (1-hidden, 71.67)};
            \addplot [fill=teal!50!white]
            	coordinates {(linear, 70.43) (linear concat, 74.98) (1-hidden, 80.17)};
            \addplot [fill=cyan!60!white,]
            	coordinates {(linear, 69.55) (linear concat, 74.70) (1-hidden, 79.63)};
            \addplot [fill=blue!75!white,]
            	coordinates {(linear, 70.75) (linear concat, 75.81) (1-hidden, 79.84)};
            \addplot [fill=red!75!white,]
            	coordinates {(linear, 72.11) (linear concat, 76.91) (1-hidden, 80.84)};
        \end{axis}
    \end{tikzpicture}
    
    \begin{tikzpicture}
        \begin{axis}[
            ybar,
            ymin = 45,
            ymax = 82,
            y label style ={at={(axis description cs:0.50,1.1)}, anchor=center, rotate=-90},
            ylabel={Phoneme classification on unseen data - TIMIT (39-classes)},
            extra y ticks = {65},
            extra y tick labels = { Accuracy (\%) },
            extra y tick style = { tick label style={rotate=90, yshift=7mm,},},
            bar width = 0.55cm,
            enlarge x limits = 0.25,
            height = 3.5cm,
            symbolic x coords = {linear, linear concat, 1-hidden},
            xtick=data,
            x=5.0cm,
            nodes near coords,
            nodes near coords align = {vertical},
            nodes near coords style = {
                font=\footnotesize,
                /pgf/number format/.cd,
                fixed,
                fixed zerofill,
                precision=1,},
            ]
            \addplot [pattern=crosshatch dots, pattern color=blue,]
            	coordinates {(linear, 50.28) (linear concat, 58.22) (1-hidden, 52.54)};
            \addplot [pattern=crosshatch dots, pattern color=red,]
            	coordinates {(linear, 63.26) (linear concat, 68.34) (1-hidden, 64.90)};
            \addplot [pattern=crosshatch dots, pattern color=orange,]
            	coordinates {(linear, 63.29) (linear concat, 68.26) (1-hidden, 65.38)};
            \addplot [fill=teal!50!white]
            	coordinates {(linear, 70.78) (linear concat, 72.22) (1-hidden, 73.33)};
            \addplot [fill=cyan!60!white,]
            	coordinates {(linear, 70.08) (linear concat, 72.26) (1-hidden, 72.45)};
            \addplot [fill=blue!75!white,]
            	coordinates {(linear, 70.52) (linear concat, 72.72) (1-hidden, 72.97)};
            \addplot [fill=red!75!white,]
            	coordinates {(linear, 71.48) (linear concat, 73.44) (1-hidden, 73.86)};
        \end{axis}
    \end{tikzpicture}

    \begin{tikzpicture}
        \begin{axis}[
            ybar,
            ymin = 50,
            ymax = 113,
            y label style ={at={(axis description cs:0.50,1.1)}, anchor=center, rotate=-90},
            ylabel={Classification on more speech downstream tasks},
            extra y ticks = {85},
            extra y tick labels = { Accuracy (\%) },
            extra y tick style = { tick label style={rotate=90, yshift=7mm,},},
            bar width = 0.55cm,
            enlarge x limits = 0.25,
            height = 3.5cm,
            legend cell align = {left},
            legend style = {at={(0.5,-0.5),}, draw=none, /tikz/every even column/.append style = {column sep=6pt}, anchor=north, legend columns=-1},
            transpose legend,
            symbolic x coords = {keyword spotting (unseen), speaker frame (seen), speaker utter (seen)},
            xtick=data,
            x=5.0cm,
            nodes near coords,
            nodes near coords align = {vertical},
            nodes near coords style = {
                font=\footnotesize,
                /pgf/number format/.cd,
                fixed,
                fixed zerofill,
                precision=1,},
            ]
            \addplot [pattern=crosshatch dots, pattern color=blue,]
            	coordinates {(keyword spotting (unseen), 82.15) (speaker frame (seen), 53.77) (speaker utter (seen), 79.28)};
            \addplot [pattern=crosshatch dots, pattern color=red,]
            	coordinates {(keyword spotting (unseen), 86.63) (speaker frame (seen), 99.83) (speaker utter (seen), 99.86)};
            \addplot [pattern=crosshatch dots, pattern color=orange,]
            	coordinates {(keyword spotting (unseen), 88.41) (speaker frame (seen), 99.86) (speaker utter (seen), 99.87)};
            \addplot [fill=teal!50!white]
            	coordinates {(keyword spotting (unseen), 91.33) (speaker frame (seen), 98.13) (speaker utter (seen), 98.98)};
            \addplot [fill=cyan!60!white,]
            	coordinates {(keyword spotting (unseen), 90.49) (speaker frame (seen), 99.86) (speaker utter (seen), 99.87)};
            \addplot [fill=blue!75!white,]
            	coordinates {(keyword spotting (unseen), 88.96) (speaker frame (seen), 99.77) (speaker utter (seen), 99.72)};
            \addplot [fill=red!75!white,]
            	coordinates {(keyword spotting (unseen), 90.46) (speaker frame (seen), 99.71) (speaker utter (seen), 99.71)};
        \legend{mag, freq, freq+mag, time, time+mag, time+freq, time+freq+mag}
        \end{axis}
    \end{tikzpicture}
    
    \caption{\small\textbf{The effect of different alterations.} We use phoneme classification, keyword spotting, and speaker recognition on both seen and unseen data to study the effect of different alteration methods. All results are \textit{base} models pre-trained on the LibriSpeech train-clean-100 subset, testing accuracy on different types of classifiers are reported. Methods with unidirectional context (without time alteration) are in dotted color, methods with bidirectional context (with time alteration) are in solid color. \vspace{-5mm}}
    \label{fig:different_alterations}
\end{figure*}
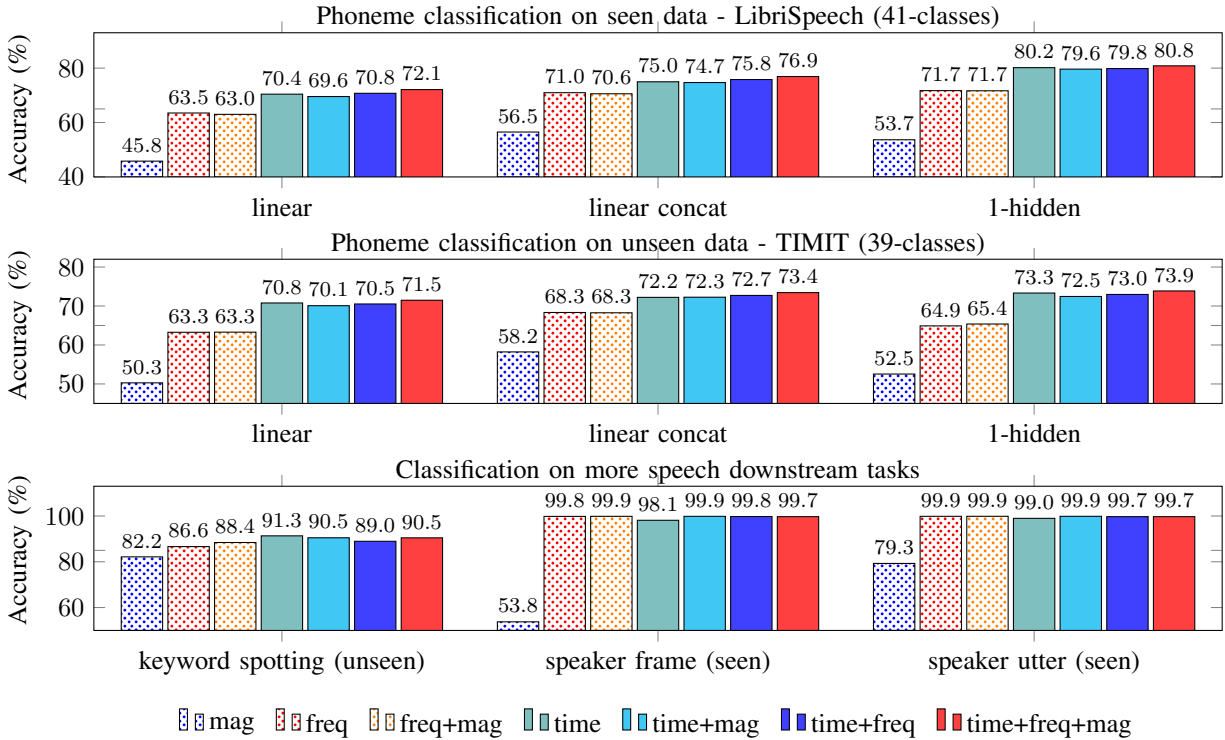

\subsection{Training Downstream Tasks}
\label{ssec:training_downstream}
For both representation extraction and fine-tuning, we use the AdamW~\cite{adamW} optimizer to update models when training with downstream tasks.
We use a learning rate of $2e^{-4}$ with a batch size of $16$.
When applying representation extraction for ASR tasks, we use the RMSPROP optimizer with a learning rate of $2e^{-4}$.
We half the learning rate every epoch if the development set error does not drop more than a threshold of $0.001$.
We use a batch size of $16$, and update for 24 epochs.
We use the same setting as above for fine-tuning on ASR, except we use a different learning rate for each TERA model.
Learning rate is set to $2e^{-4}$ when we fine-tune \textit{base}, $1e^{-4}$ for \textit{medium}, and $5e^{-5}$ for \textit{large}.
Empirically, we find that larger models require a lower learning rate during fine-tuning.
Hence, we half the learning rate every time the model depth is doubled; this helps stabilize the entire fine-tuning process.

During downstream fine-tuning, we also apply the SpecAugment~\cite{spec_augment} LD policy.
During SpecAugment, the chosen (and possibly overlapping) time and frequency blocks are zeroed out.
We omit the use of time warping as masking provides enough regularization.
We find SpecAugment is additive to the proposed pre-training approach, as it delays overfitting and improves the final accuracy numbers in the downstream tasks.

\section{Results}
\label{sec:results}
In Section~\ref{ssec:exp_different_alterations}, we study the effect of applying different combinations of time, frequency, and magnitude alteration.
In Section~\ref{ssec:exp_benchmark}, we compare TERA with recent self-supervised representation methods on downstream tasks.
In Section~\ref{ssec:exp_analysis}, we study multiple aspects of the TERA pre-training process, including fine-tuning with downstream models, the amount of unlabeled data, the effect of pre-training on various features, various model sizes, and different masking policies.
In Section~\ref{ssec:exp_asr_representation}, we compare TERA with approaches where we train ASR models on frozen speech representations.
In Section~\ref{ssec:exp_asr_finetune}, TERA is compared with approaches that fine-tune their pre-trained models.
In Section~\ref{ssec:exp_timit}, we demonstrate transferring TERA representations from one dataset to another by presenting ASR PER on TIMIT~\cite{timit}.

\subsection{The Effect of Different Alterations}
\label{ssec:exp_different_alterations}
With three types of alterations, there are a total of seven combinations, namely \textit{"time"}, \textit{"freq"}, \textit{"mag"}, \textit{"time+freq"}, \textit{"time+mag"}, \textit{"freq+mag"}, and \textit{"time+freq+mag"}.
We pre-train TERA with all seven combinations of alterations on 100 hours of LibriSpeech, then measure the learned representations with downstream tasks.
The results are presented in Figure~\ref{fig:different_alterations}.
We use TERA for representation extraction (from the last layer), i.e., TERA parameters were frozen when adapting downstream models.
We split our methods into two categories: one with the time alteration that allows the model to encode bidirectional information (denoted in a solid color).
The second category is the models that don't include the time alteration method (denoted in dotted color).
The conclusions made in the following sections are statistically significant, where we measured p-values with statistical significance tests (paired samples t-test or Fisher’s exact test). Overall, applying all alterations \textit{"time+freq+mag"} achieves the best performance on average (with an averaged p-value of 0.0492 across nine tasks when compared to \textit{"time"}).
Now, we discuss the effect of each alteration in detail.

\begin{table*}[ht]
\centering
\begin{tabular}{|r||c|c|c|c|c|}
\hline
Representation & Network & \#Params & Input Feature & Pre-train & Learning Style \\ \hline \hline
MFCC & - & 0 & 13-dim + delta 2 & - & - \\
FBANK & - & 0 & 80-dim + delta 2 & - & - \\
log Mel & - & 0 & 80-dim & - & -\\ \hline
APC~\cite{apc1, apc2} & 3-GRU & 4,064,256 & 80-dim log Mel & LibriSpeech 360 hr & autoregressive reconstruction\\
VQ-APC~\cite{vq_apc} & 3-GRU & 4,064,256 & 80-dim log Mel & LibriSpeech 360 hr & autoregressive reconstruction + VQ \\ \hline
CPC~\cite{cpc, modified_cpc} & 5-Conv 1-Trans  & 1,843,456 & waveform & LibriLight 60k hr & contrastive \\
vq-wav2vec~\cite{vq_wav2vec} & 20-Conv & 6,042,624 & waveform & LibriSpeech 960 hr & contrastive + VQ \\
wav2vec 2.0~\cite{wav2vec2} & 7-Conv 24-Trans & 214,658,176 & waveform & LibriSpeech 960 hr & contrastive + latent masking \\ \hline
Mockingjay~\cite{mockingjay} & 3-Trans & 21,327,360 & 80-dim log Mel & LibriSpeech 960 hr & time-only masked reconstruction \\
Audio ALBERT~\cite{audioalbert} & 3-Trans & 7,151,616 & 80-dim log Mel & LibriSpeech 960 hr & time-only masked reconstruction \\
NPC~\cite{npc} & 3-Conv & 19,273,728 & 80-dim log Mel & LibriSpeech 360 hr & time-only masked reconstruction \\  \hline
TERA (Ours) & 3-Trans & 21,327,360 & 80-dim log Mel & LibriSpeech 960 hr & "time+freq+mag" alteration \\ \hline

\end{tabular}
\caption{\small\textbf{Details of baseline features and recent speech representation approaches.} We evaluate their performance on downstream tasks and present the results in Table~\ref{table:benchmark}. \vspace{-5mm} } 
\label{table:upstreams}
\end{table*}

\subsubsection{The Effect of Time Alteration}
\label{sssec:exp_time}
We discuss the effect of time alteration for each downstream task.
For phoneme classification, the models with time alteration (in solid color) perform reasonably well, while the models without time alteration (in dotted color) perform relatively poorly.
The results of seen data (LibriSpeech) and unseen data (TIMIT) agree with each other and yield similar trends.
For keyword spotting, the \textit{"time"} model achieves the best performance.
Adding other alterations compromise the performance.
The models with time alteration perform better than the models that do not have time alteration.
This observation suggests that bidirectional understanding is an essential aspect of detection tasks.
For speaker recognition, the existence of time alteration does not compromise the performance.
With time alteration alone, the \textit{"time"} model achieves slightly lower but satisfactory performance.
From the above discussion, we thus deem the time alteration to be indispensable.
We speculate that although Transformer Encoders~\cite{transformer} are bidirectional due to their multi-head self-attention, without the time alteration, the network fails to encode proper context and yield sub-optimal performance.
However, through alteration on the time axis, the model establishes a bidirectional understanding of the audio and gives better results.

\begin{table*}[ht]
\centering
\begin{tabular}{|r||ccc|ccc|c|cc|c|}
\hline
\multicolumn{1}{|r||}{\multirow{2}{*}{Representation}} & \multicolumn{3}{c|}{Phoneme - LibriSpeech} & \multicolumn{3}{c|}{Phoneme - TIMIT} & Keyword & \multicolumn{2}{c|}{Speaker} & \multirow{2}{*}{Average} \\
\multicolumn{1}{|l||}{} & linear & linear concat & 1-hidden & linear & linear concat & 1-hidden & spotting & frame & utter &  \\
\hline \hline
MFCC & 47.88 & 51.46 & 62.47 & 54.74 & 58.55 & 65.52 & 87.99 & 13.18 & 90.61 & 59.16 \\
FBANK & 48.01 & 52.16 & 63.38 & 55.00 & 58.35 & 64.52 & 86.01 & 31.35 & 94.70 & 61.50 \\
log Mel & 41.93 & 49.94 & 48.55 & 47.37 & 56.04 & 51.05 & 72.35 & 22.38 & 95.33 & 53.88 \\ \hline
APC~\cite{apc1, apc2} & 72.76 & 79.62 & 78.06 & 72.90 & 76.74 & 74.51 & 91.04 & 79.08 & \textbf{99.63} & 80.48 \\
VQ-APC~\cite{vq_apc} & 71.92 & 79.08 & 77.62 & 72.06 & 75.98 & 74.11 & 92.92 & 68.56 & 99.21 & 79.05 \\ \hline
CPC~\cite{cpc, modified_cpc} & 71.31 & 78.52 & 77.40 & 73.10 & \textbf{77.71} & 75.57 & 94.16 & 75.47 & 99.32 & 80.28 \\
vq-wav2vec~\cite{vq_wav2vec} & 63.38 & 76.53 & 66.18 & 67.52 & 76.42 & 68.68 & \textbf{94.55} & 24.76 & 84.09 & 69.12 \\
wav2vec 2.0~\cite{wav2vec2} & 56.39 & 72.72 & 73.61 & 59.32 & 71.76 & 69.14 & 68.87 & 82.53 & 95.40 & 72.19 \\ \hline
Mockingjay~\cite{mockingjay} & 67.51 & 72.07 & 78.70 & 69.21 & 71.68 & 72.53 & 88.09 & 97.22 & 98.35 & 79.48 \\
Audio ALBERT~\cite{audioalbert} & 68.13 & 72.62 & 78.65 & 69.62 & 71.37 & 72.68 & 90.59 & 96.65 & 98.37 & 79.85 \\
NPC~\cite{npc} & 67.32 & 76.08 & 75.35 & 65.76 & 72.96 & 68.42 & 87.41 & 27.72 & 96.16 & 70.80 \\ \hline
TERA time+freq & \textbf{74.45} & \textbf{78.84} & \textbf{82.14} & 73.92 & 75.80 & 75.98 & 91.89 & \textbf{99.57} & 99.59 & 83.58 \\
+mag & 74.07 & 78.67 & 81.90 & \textbf{74.14} & 75.92 & \textbf{76.23} & 92.60 & 99.47 & 99.48 & \textbf{83.61} \\ \hline
\end{tabular}
\caption{\small\textbf{Performance of baseline features and recent speech representation approaches.} Representations listed in Table~\ref{table:upstreams} are evaluated on the four downstream tasks. We report testing accuracy (\%) and highlight the highest score for each column in bold font.} 
\label{table:benchmark}
\end{table*}

\subsubsection{The Effect of Frequency Alteration}
\label{sssec:exp_freq}
For the phoneme classification tasks, adding the frequency alteration during pre-training is effective in boosting performance.
The \textit{"time+freq"} and \textit{"time+freq+mag"} model improves over the \textit{"time"} model for both seen and unseen data.
For keyword spotting, adding the frequency alteration does not help.
For speaker recognition, we find that the model can achieve strong speaker classification performance by employing the frequency alteration.
The \textit{"freq"} model, with frequency alteration alone, is enough to achieve high performance.
The \textit{"mag"} and \textit{"time"} models, without frequency alteration, have lower performance.
By employing the frequency alteration, models benefit in the phoneme classification and speaker recognition performance, which is surprising and counter-intuitive. 
It is well known that a robust phonetic system should be speaker invariant~\cite{vc}.
We surmise this phenomenon results from that TERA representations provide more accessibility to phonetic and speaker information and maintain the separability between the two types of information.
Downstream models can efficiently learn to extract task-specific information.
To sum up this section, the frequency alteration effectively encodes speaker identity while also fine grinding the phonetic quality.

\subsubsection{The Effect of Magnitude Alteration}
\label{sssec:exp_mag}
For the phoneme classification tasks, adding the magnitude alteration is effective in boosting performance.
The \textit{"time+freq+mag"} model improves over the \textit{"time+freq"} model for both seen and unseen data.
For keyword spotting, adding the magnitude alteration improves most cases but does not improve over time alteration alone.
For speaker recognition, adding the magnitude alteration improves performance for all cases.
We observe that by applying the magnitude alteration during pre-training, the model learns to be robust to various inputs.
As a result, we improve downstream tasks performance by adding the magnitude alteration.

\subsection{Comparison of Recent Speech Representation Approaches}
\label{ssec:exp_benchmark}
In this section, we compare TERA with recent speech representation learning methods through evaluating with downstream tasks.
We select eight different methods, which cover a wide range of representation learning techniques.
We list these methods and their details in Table~\ref{table:upstreams}.
We organize Table~\ref{table:upstreams} by first grouping methods with similar learning styles, then by their publication date in the order of first to last.
We use publicly available pre-trained weights or pre-training scripts provided by their original authors for all of the listed models.
Note that all models are pre-trained on LibriSpeech~\cite{librispeech} except for CPC~\cite{cpc, modified_cpc}, which is pre-trained on LibriLight~\cite{librilight}.
All of the models and baseline features have a stride of 10 ms, with only one exception where wav2vec 2.0~\cite{wav2vec2} has a downsample rate of 320, which generates a representation every 20 ms.
Interestingly, all of the methods that employ reconstruction loss in their learning objective use an 80-dim log Mel input feature.
We also use 80-dim log Mel for consistency and fair comparison.
On the other hand, all methods that employ contrastive loss in their learning objective use raw waveform as an input feature.
In Table~\ref{table:upstreams}, we also list the details of baseline features.
We present the performance of different representations on downstream tasks in Table~\ref{table:benchmark}.
The pre-trained models are all used for representation extraction (from the last layer), i.e., we freeze parameters when adopting the downstream model and feed representations to the downstream model as input.
In the last column, we average the accuracy of all tasks across each row.
The above results can be reproduced with the S3PRL toolkit\textsuperscript{\ref{github}}, where all the self-supervised models and downstream tasks are available with easy-to-use setups.

\begin{table*}[ht]
\centering
\begin{tabular}{|r||ccc|ccc|c|cc|c|}
\hline
\multicolumn{1}{|r||}{\multirow{2}{*}{Method}} & \multicolumn{3}{c|}{Phoneme - LibriSpeech} & \multicolumn{3}{c|}{Phoneme - TIMIT} & Keyword & \multicolumn{2}{c|}{Speaker} & \multirow{2}{*}{Average} \\
\multicolumn{1}{|l||}{} & linear & linear concat & 1-hidden & linear & linear concat & 1-hidden & spotting & frame & utter &  \\
\hline \hline
TERA & 74.07 & 78.67 & 81.90 & 74.14 & 75.92 & 76.23 & 92.60 & 99.47 & 99.48 & 83.61 \\
+ \textit{fine-tune} & 89.09 & 89.07 & 89.53 & 78.81 & \textbf{78.86} & 78.37 & \textbf{94.03} & 99.50 & 99.62 & 88.54 \\
+ \textit{SpecAug} & \textbf{90.36} & \textbf{90.07} & \textbf{90.88} & \textbf{79.50} & 78.69 & \textbf{78.82} & 93.74 & \textbf{99.81} & \textbf{99.86} & \textbf{89.08} \\ \hline
random init + \textit{SpecAug} & 88.70 & 89.49 & 89.08 & 70.10 & 68.09 & 71.57 & 8.34 & 0.424 & 1.67 & 54.16 \\ \hline
\end{tabular}
\caption{\small\textbf{Performance of fine-tuning TERA and baseline approaches.} Here we use the TERA-base \textit{"time+freq+mag"} model, row one is identical to the last row of Table~\ref{table:benchmark}. We report testing accuracy (\%) and highlight the highest score for each column in bold font. \vspace{-5mm} } 
\label{table:fine-tune}
\end{table*}

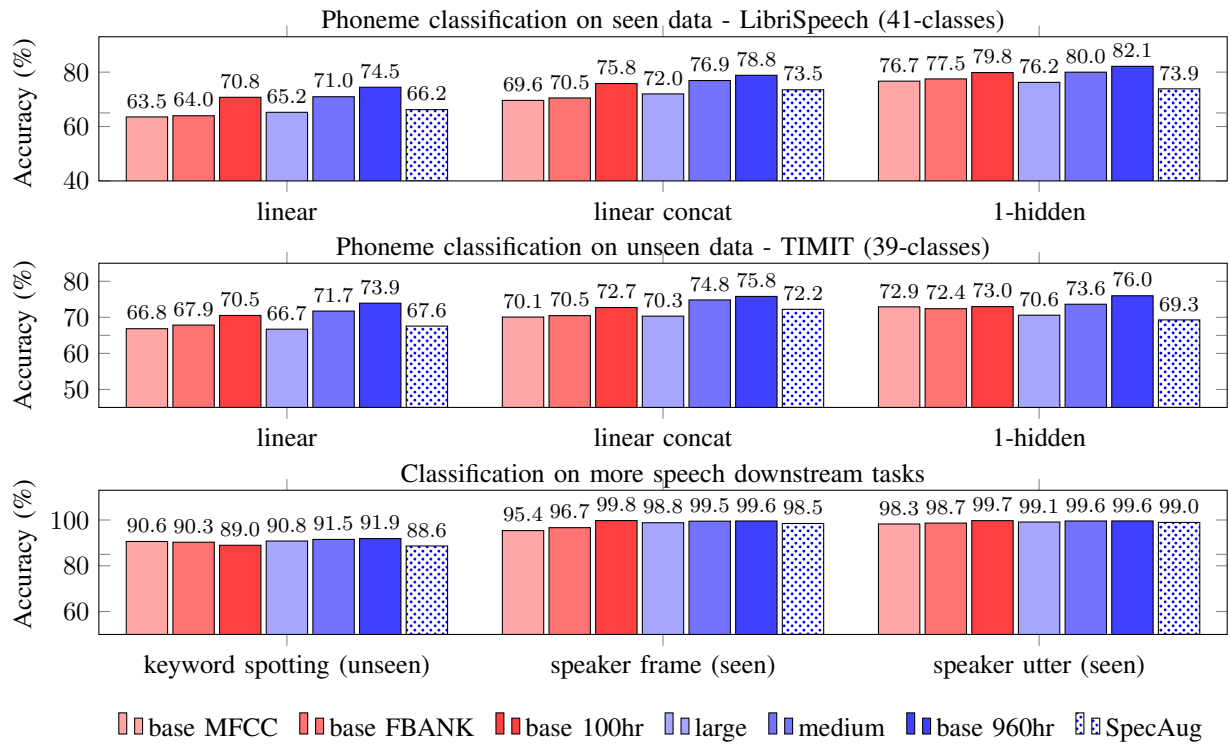
\begin{figure*}[ht]
    \centering
    
    \begin{tikzpicture}
        \begin{axis}[
            ybar,
            ymin = 40,
            ymax = 93,
            y label style ={at={(axis description cs:0.50,1.1)}, anchor=center, rotate=-90},
            ylabel={Phoneme classification on seen data - LibriSpeech (41-classes)},
            extra y ticks = {65},
            extra y tick labels = { Accuracy (\%) },
            extra y tick style = { tick label style={rotate=90, yshift=7mm,},},
            bar width = 0.55cm,
            enlarge x limits = 0.25,
            height = 3.5cm,
            symbolic x coords = {linear, linear concat, 1-hidden},
            xtick=data,
            x=5.0cm,
            nodes near coords,
            nodes near coords align = {vertical},
            nodes near coords style = {
                font=\footnotesize,
                /pgf/number format/.cd,
                fixed,
                fixed zerofill,
                precision=1,},
            ]
            \addplot [fill=red!35!white,]
            	coordinates {(linear, 63.53) (linear concat, 69.62) (1-hidden, 76.68)};
            \addplot [fill=red!55!white,]
            	coordinates {(linear, 63.96) (linear concat, 70.50) (1-hidden, 77.50)};
            \addplot [fill=red!75!white,]
            	coordinates {(linear, 70.75) (linear concat, 75.81) (1-hidden, 79.84)};
            \addplot [fill=blue!35!white,]
            	coordinates {(linear, 65.23) (linear concat, 71.99) (1-hidden, 76.23)};
            \addplot [fill=blue!55!white,]
            	coordinates {(linear, 70.95) (linear concat, 76.91) (1-hidden, 79.98)};
            \addplot [fill=blue!75!white,]
            	coordinates {(linear, 74.45) (linear concat, 78.84) (1-hidden, 82.14)};
            \addplot [pattern=crosshatch dots, pattern color=blue,]
            	coordinates {(linear, 66.21) (linear concat, 73.50) (1-hidden, 73.85)};
        \end{axis}
    \end{tikzpicture}
    
    \begin{tikzpicture}
        \begin{axis}[
            ybar,
            ymin = 45,
            ymax = 85,
            y label style ={at={(axis description cs:0.50,1.1)}, anchor=center, rotate=-90},
            ylabel={Phoneme classification on unseen data - TIMIT (39-classes)},
            extra y ticks = {65},
            extra y tick labels = { Accuracy (\%) },
            extra y tick style = { tick label style={rotate=90, yshift=7mm,},},
            bar width = 0.55cm,
            enlarge x limits = 0.25,
            height = 3.5cm,
            symbolic x coords = {linear, linear concat, 1-hidden},
            xtick=data,
            x=5.0cm,
            nodes near coords,
            nodes near coords align = {vertical},
            nodes near coords style = {
                font=\footnotesize,
                /pgf/number format/.cd,
                fixed,
                fixed zerofill,
                precision=1,},
            ]
            \addplot [fill=red!35!white,]
            	coordinates {(linear, 66.84) (linear concat, 70.07) (1-hidden, 72.90)};
            \addplot [fill=red!55!white,]
            	coordinates {(linear, 67.85) (linear concat, 70.46) (1-hidden, 72.39)};
            \addplot [fill=red!75!white,]
            	coordinates {(linear, 70.52) (linear concat, 72.72) (1-hidden, 72.97)};
            \addplot [fill=blue!35!white,]
            	coordinates {(linear, 66.71) (linear concat, 70.33) (1-hidden, 70.59)};
            \addplot [fill=blue!55!white,]
            	coordinates {(linear, 71.73) (linear concat, 74.82) (1-hidden, 73.63)};
            \addplot [fill=blue!75!white,]
            	coordinates {(linear, 73.92) (linear concat, 75.80) (1-hidden, 75.98)};
            \addplot [pattern=crosshatch dots, pattern color=blue,]
            	coordinates {(linear, 67.56) (linear concat, 72.22) (1-hidden, 69.28)};
        \end{axis}
    \end{tikzpicture}

    \begin{tikzpicture}
        \begin{axis}[
            ybar,
            ymin = 50,
            ymax = 113,
            y label style ={at={(axis description cs:0.50,1.1)}, anchor=center, rotate=-90},
            ylabel={Classification on more speech downstream tasks},
            extra y ticks = {85},
            extra y tick labels = { Accuracy (\%) },
            extra y tick style = { tick label style={rotate=90, yshift=7mm,},},
            bar width = 0.55cm,
            enlarge x limits = 0.25,
            height = 3.5cm,
            legend cell align = {left},
            legend style = {at={(0.5,-0.5),}, draw=none, /tikz/every even column/.append style = {column sep=6pt}, anchor=north, legend columns=-1},
            transpose legend,
            symbolic x coords = {keyword spotting (unseen), speaker frame (seen), speaker utter (seen)},
            xtick=data,
            x=5.0cm,
            nodes near coords,
            nodes near coords align = {vertical},
            nodes near coords style = {
                font=\footnotesize,
                /pgf/number format/.cd,
                fixed,
                fixed zerofill,
                precision=1,},
            ]
            \addplot [fill=red!35!white,]
            	coordinates {(keyword spotting (unseen), 90.62) (speaker frame (seen), 95.37) (speaker utter (seen), 98.26)};
            \addplot [fill=red!55!white,]
            	coordinates {(keyword spotting (unseen), 90.30) (speaker frame (seen), 96.65) (speaker utter (seen), 98.65)};
            \addplot [fill=red!75!white,]
            	coordinates {(keyword spotting (unseen), 88.96) (speaker frame (seen), 99.77) (speaker utter (seen), 99.72)};
            \addplot [fill=blue!35!white,]
            	coordinates {(keyword spotting (unseen), 90.81) (speaker frame (seen), 98.80) (speaker utter (seen), 99.12)};
            \addplot [fill=blue!55!white,]
            	coordinates {(keyword spotting (unseen), 91.53) (speaker frame (seen), 99.49) (speaker utter (seen), 99.57)};
            \addplot [fill=blue!75!white,]
            	coordinates {(keyword spotting (unseen), 91.89) (speaker frame (seen), 99.57) (speaker utter (seen), 99.59)};
            \addplot [pattern=crosshatch dots, pattern color=blue,]
            	coordinates {(keyword spotting (unseen), 88.64) (speaker frame (seen), 98.49) (speaker utter (seen), 98.95)};
        \legend{base MFCC, base FBANK, base 100hr, large, medium, base 960hr, SpecAug}
        \end{axis}
    \end{tikzpicture}
    
    \caption{\small\textbf{Analysis on TERA.} We pre-train TERA with different amount of unlabeled data, with different features, and with different model depth. Moreover, we directly apply the SpecAugment~\cite{spec_augment} LD Policy for pre-training instead of TERA's alteration, we get performance degrade. The models trained with 100 hours are denoted in \textcolor{red}{red}, and the models trained with 960 hours are in denoted \textcolor{blue}{blue}. \vspace{-5mm}}
    \label{fig:analysis}
\end{figure*}

\subsubsection{Phoneme Classification Results}
\label{ssec:exp_benchmark-phoneme}
For phoneme classification on seen data (LibriSpeech), TERA outperforms other methods.
For phoneme classification on unseen data (TIMIT), TERA outperforms other methods except for CPC on the \textit{linear concat} classifier.
Some representations encounter a more significant degradation in performance for unseen data.
On the other hand, TERA is not affected by the domain change.
In particular, we find that adding the TERA magnitude alteration helps performance for unseen data.
As a result, by adding magnitude alteration, we increase accuracy for phoneme classification on TIMIT.
We observe that all representations benefit from concatenating neighboring frames, as their performance on \textit{linear concat} is better than \textit{linear}.
The \textit{linear concat} classifier provided neighboring information through the concatenated frames.
Several representations benefit more from \textit{linear concat} than others, depending on how much temporal information the single feature frame already encodes.
We also observe that all representations benefit from a deeper downstream model, as their performance on \textit{1-hidden} is better than \textit{linear}.
The \textit{1-hidden} classifier is able to extract more information than linear classifiers.
The wav2vec 2.0 method achieves the lowest accuracy for phoneme classification, which contrasts its high performance when used as an initialization for ASR encoder fine-tuning~\cite{wav2vec2}.
The under-performing of wav2vec 2.0 here suggests that although large and deep models are suitable for downstream initialization and fine-tuning, they may not be a good choice for feature extraction.
Our later experiment shows that TERA-large under-perform TERA-base for feature extraction, but TERA-large outperforms TERA-base for ASR fine-tuning.

\subsubsection{Keyword Spotting Results}
\label{ssec:exp_benchmark-keyword}
For keyword spotting, all representations achieve a good score.
However, wav2vec 2.0 and NPC are not able to surpass the performance of baseline features.
The wav2vec 2.0 method achieves the lowest score, which again suggests that pre-trained models may work well for fine-tuning, but it does not always imply that they can generate meaningful representations.
The NPC method and Mockingjay achieve comparable performance, where Mockingjay barely exceeds the performance of MFCC.
By adding frequency alteration and magnitude alteration to Mockingjay, TERA can boost the performance by 4.51\% over Mockingjay.
The best performing representation on keyword spotting is vq-wav2vec, which is surprising as it does not perform as well as the others on phoneme classification and speaker recognition.
CPC achieves comparable performance with vq-wav2vec, suggesting that contrastive learning over time may be the key for detection tasks.
Other than CPC and vq-wav2vec, VQ-APC and TERA achieve high scores on this task too.

\subsubsection{Speaker Recognition Results}
\label{ssec:exp_benchmark-speaker}
For the speaker recognition task, TERA achieves the highest speaker classification accuracy for both the \textit{frame} and \textit{utter} setting.
We can observe a discriminative comparison under the \textit{frame} setting.
Methods that use autoregressive, contrastive, vector-quantization have a lower frame-wise speaker recognition accuracy.
The autoregressive prediction method focuses more on the local dependencies between time steps. Also, contrastive learning discriminates input from a distant time, and vector-quantization is a bottleneck that limits information flow over the network.
The above techniques seem to have encouraged the model not to encode speaker information in each frame, resulting in poor \textit{frame} classification performance.
Although the NPC technique uses time-masked reconstruction like Mockingjay and Audio ALBERT, it performs poorly for the \textit{frame} setting.
We speculate that this is because Masked Convolution Blocks' masking is fixed and designed to focus on local dependencies.
As a result, each representation doesn't preserve speaker information.
On the other hand, the time mask of Mockingjay, Audio ALBERT, and TERA is applied through dynamic masking~\cite{mockingjay, audioalbert}, where masks may overlap each other, creating various time mask lengths.
The overlapped masks encourage the model to also focus on global information.
As speaker characteristics tend to persist across time, this thus preserves the speaker information.
For the \textit{utter} setting of speaker recognition, all of the representations yield satisfactory results except for vq-wav2vec.
However, remember that vq-wav2vec achieves the highest score on keyword spotting.
We conclude that there is a trade-off for some representation learning methods, vq-wav2vec is good on keyword spotting but lacks speaker information.

\subsection{Analysis on TERA}
\label{ssec:exp_analysis}
To better understand how TERA derives representation from speech, we study various aspects of TERA's pre-training.
We first investigate how TERA can benefit from fine-tuning, where we present results in Table~\ref{table:fine-tune}.
Next, we investigate the effect of increasing the amount of pre-training data.
Also, we study the effect of learning with various acoustic features.
Furthermore, we investigate how the network depth (number of layers) affects TERA's performance.
Finally, we study the difference between directly applying SpecAugment and the TERA time and frequency alteration.
We intentionally did not apply the magnitude alteration here to rule out the variation.
We visualize results in Figure~\ref{fig:analysis}.
We denote models trained on 100 hours of LibriSpeech in red and denote models trained on 960 hours of LibriSpeech in blue.
All TERA models are pre-trained with time and frequency alteration.
We freeze all models for representation extraction from the last layer.

\subsubsection{Fine-tuning TERA}
\label{ssec:exp_fine_tune}
In Table~\ref{table:fine-tune}, we show the effect of fine-tuning the pre-trained TERA with downstream tasks.
In the last column, we average the accuracy of all tasks across each row.
We fine-tune TERA \textit{"time+freq+mag"}, the best performing TERA-base model, which is pre-trained on 960 hours of LibriSpeech.
In the first row of Table~\ref{table:fine-tune}, we include the results from the last row of Table~\ref{table:benchmark} of the frozen TERA \textit{"time+freq+mag"} model to link the two tables.
With fine-tuning, we see considerable performance increases for all tasks.
By adding SpecAugment~\cite{spec_augment} during fine-tune, we improve performance for some tasks.
However, the improvement of adding SpecAugment during fine-tuning is limited because the pre-trained weights are good initialization for the tasks.
Empirically, we find that when compared to training downstream classifiers on frozen TERA, fine-tuning TERA with downstream tasks dramatically reduces the number of training steps for the downstream models to converge.
We also observe that fine-tuning for phoneme classification on LibriSpeech brings a more significant improvement than on TIMIT.
The reason is that there are more labeled data for LibriSpeech.
It is worth noticing that fine-tuning with limited label data (TIMIT) still benefits performance, where we observe no sign of overfitting.

We train the same TERA network (with randomly initialized parameters) on downstream tasks, where we distort the input features with SpecAugment to prevent overfitting.
The above setting serves as a baseline for fine-tuning pre-trained TERA, and we denote it as \textit{random init + SpecAug} in Table~\ref{table:fine-tune}.
Overall, the downstream models trained with randomly initialized TERA and SpecAugment either underperform or perform poorly.
The baseline \textit{random init + SpecAug} results of phoneme classification on LibriSpeech are close to the pre-trained models because the amount of label is sufficient.
However, there is a performance gap between baseline results and pre-trained models on phoneme classification for TIMIT.
The reason is the limited amount of label data.
The baseline \textit{random init + SpecAug} overfits for the keyword spotting and speaker recognition task. 
Note that the reported accuracy is from the test set; therefore, a low testing score means severe overfitting on the training set.
It does not indicate that the network is untrainable.
We conclude that without pre-training, model architecture alone provides no benefit.

\begin{figure*}[ht]
    \centering
    
    \begin{tikzpicture}
        \begin{axis}[
            ybar,
            ymin = 26,
            ymax = 117,
            y label style ={at={(axis description cs:0.50,1.1)}, anchor=center, rotate=-90},
            ylabel={Classification on seen data - LibriSpeech},
            extra y ticks = {75},
            extra y tick labels = { Accuracy (\%) },
            extra y tick style = { tick label style={rotate=90, yshift=7mm,},},
            bar width = 0.40cm,
            enlarge x limits = 0.13,
            height = 3.5cm,
            symbolic x coords = {linear, linear concat, 1-hidden, keyword spotting},
            symbolic x coords = {linear, linear concat, 1-hidden, speaker frame, speaker utter},
            xtick=data,
            x=3.2cm,
            nodes near coords,
            nodes near coords align = {vertical},
            nodes near coords style = {
                font=\footnotesize,
                /pgf/number format/.cd,
                fixed,
                fixed zerofill,
                precision=0,},
            ]
            \addplot [pattern=crosshatch dots, pattern color=orange,]
            	coordinates {(linear, 67.32) (linear concat, 76.08) (1-hidden, 75.35) (speaker frame, 27.72) (speaker utter, 96.16)};
            \addplot [fill=orange!55!white,]
            	coordinates {(linear, 66.33) (linear concat, 74.59) (1-hidden, 74.31) (speaker frame, 27.66) (speaker utter, 95.19)};
            \addplot [pattern=crosshatch dots, pattern color=red,]
            	coordinates {(linear, 70.43) (linear concat, 74.98) (1-hidden, 79.28) (speaker frame, 98.13) (speaker utter, 98.98)};
            \addplot [fill=red!55!white]
            	coordinates {(linear, 67.51) (linear concat, 72.07) (1-hidden, 78.70) (speaker frame, 97.22) (speaker utter, 98.35)};
            \addplot [pattern=crosshatch dots, pattern color=cyan,]
            	coordinates {(linear, 70.75) (linear concat, 75.81) (1-hidden, 79.84) (speaker frame, 99.77) (speaker utter, 99.72)};
            \addplot [fill=cyan!65!white,]
            	coordinates {(linear, 74.45) (linear concat, 78.84) (1-hidden, 82.14) (speaker frame, 99.57) (speaker utter, 99.59)};
        \end{axis}
    \end{tikzpicture}

    \begin{tikzpicture}
        \begin{axis}[
            ybar,
            ymin = 60,
            ymax = 103,
            y label style ={at={(axis description cs:0.5,1.1)}, anchor=center, rotate=-90},
            ylabel={Classification on unseen data - TIMIT / Speech Commands},
            extra y ticks = {80},
            extra y tick labels = { Accuracy (\%) },
            extra y tick style = { tick label style={rotate=90, yshift=7mm,},},
            bar width = 0.40cm,
            enlarge x limits = 0.2,
            height = 3.5cm,
            legend cell align = {left},
            legend style = {at={(0.5,-0.5),}, draw=none, /tikz/every even column/.append style = {column sep=6pt}, anchor=north, legend columns=-1},
            transpose legend,
            symbolic x coords = {linear, linear concat, 1-hidden, keyword spotting},
            xtick=data,
            x=3.3cm,
            nodes near coords,
            nodes near coords align = {vertical},
            nodes near coords style = {
                font=\footnotesize,
                /pgf/number format/.cd,
                fixed,
                fixed zerofill,
                precision=0,},
            ]
            \addplot [pattern=crosshatch dots, pattern color=orange,]
            	coordinates {(linear, 65.76) (linear concat, 72.96) (1-hidden, 68.42) (keyword spotting, 87.41)};
            \addplot [fill=orange!55!white,]
            	coordinates {(linear, 64.78) (linear concat, 73.34) (1-hidden, 67.86) (keyword spotting, 93.25)};
            \addplot [pattern=crosshatch dots, pattern color=red,]
            	coordinates {(linear, 70.78) (linear concat, 72.22) (1-hidden, 73.33) (keyword spotting, 91.33)};
            \addplot [fill=red!55!white]
            	coordinates {(linear, 69.21) (linear concat, 71.68) (1-hidden, 72.53) (keyword spotting, 88.09)};
            \addplot [pattern=crosshatch dots, pattern color=cyan,]
            	coordinates {(linear, 70.52) (linear concat, 72.72) (1-hidden, 72.97) (keyword spotting, 88.96)};
            \addplot [fill=cyan!65!white,]
            	coordinates {(linear, 73.92) (linear concat, 75.80) (1-hidden, 75.98) (keyword spotting, 91.89)};
        \legend{NPC 360hr, NPC 960hr, Mockingjay 100hr, Mockingjay 960hr, TERA 100hr, TERA 960hr}
        \end{axis}
    \end{tikzpicture}
    
    \caption{\small\textbf{Pre-training on more data.} We pre-train different masked reconstruction models on a different amount of unlabeled data. We observe that NPC and Mockingjay got worse performance as we add noisy data (LibriSpeech \textit{train-other-500}) during pre-train, mainly due to the reconstruction nature that remembers everything. TERA, on the other hand, does not suffer from this issue. \vspace{-5mm}}
    \label{fig:masked_reconstruction}
\end{figure*}
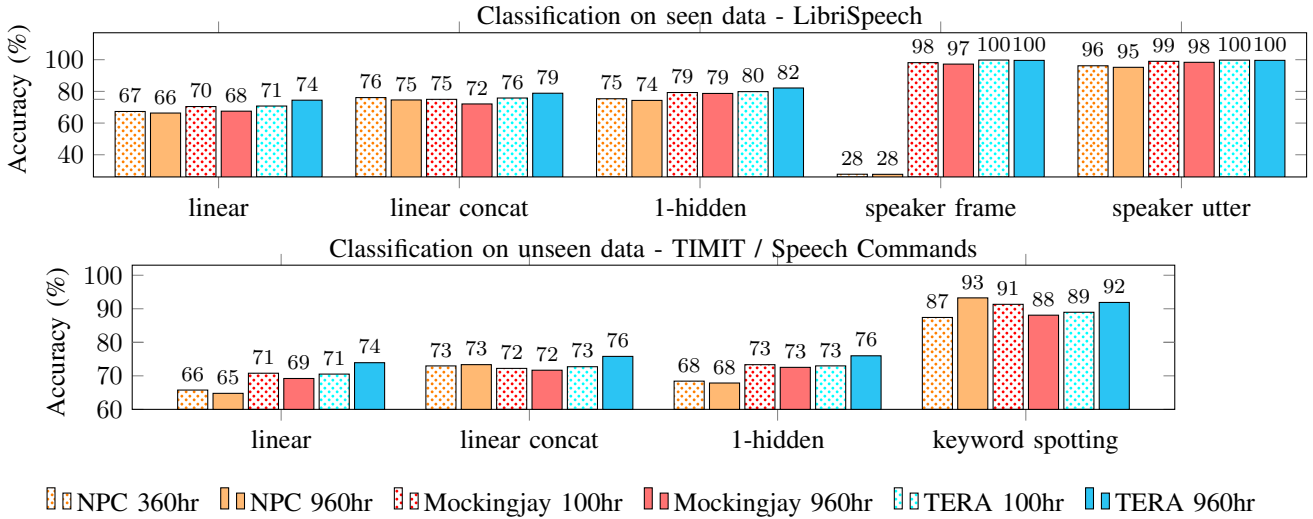

\subsubsection{Pre-training on More Data}
\label{ssec:exp_analysis_more_data}
Unsurprisingly, pre-training TERA on more data increases the performance of all downstream tasks.
In Figure~\ref{fig:analysis}, \textit{base 960hr} improves significantly over \textit{base 100hr}.
Presumably, all pre-trained models should benefit from more unlabeled data.
However, we find that this is not the case for all methods.
In particular, we find that models trained from time-only masked reconstruction (See Table~\ref{table:upstreams}) may not improve by training on more data, especially when the added data is unclean.
To be more specific, we find that Mockingjay~\cite{mockingjay} and NPC~\cite{npc} got worse performance when we add \textit{train-other-500} to the pre-training data.
The Mockingjay method uses dynamic masking on data along the time-axis.
The NPC method uses a mask in its convolution module to achieve masking on the time-axis.
These methods use the idea of reconstructing a masked time span from surrounding context, which is unlike other pre-training methods such as the line of work of APC~\cite{apc1, apc2, vq_apc, improved_apc} that uses autoregressive prediction and the line of work of CPC~\cite{cpc, modified_cpc, bidir_cpc, wav2vec, vq_wav2vec, vq_wav2vec_ft, wav2vec2} that uses contrastive learning.
The time-only masked reconstruction process forces the model to remember all the information of speech, including speaker characteristics and other noise.
However, with the regularization of frequency masking, the TERA representations do not exhibit this weakness.

This observation is presented in Figure~\ref{fig:masked_reconstruction}, where NPC~\cite{npc}, Mockingjay~\cite{mockingjay}, and TERA \textit{"time+freq"} are pre-trained with increasing amounts of unlabeled data. 
We evaluate the learned representations on downstream tasks.
For all tasks and all types of classifiers, both NPC and Mockingjay got worse performance when we add the \textit{train-other-500} subset for pre-train.
There is only one exception with NPC, where it improves on the keyword spotting task by pre-training on more data.
On the other hand, TERA can benefit from training on noisy data.
As a result, TERA improves for all tasks by pre-training on more data.

We want to point out that pre-training on more data has not been explored in NPC and Mockingjay, where they only report results with models pre-trained on 360 hours of LibriSpeech.
We believe this is a unique phenomenon for methods that learn from the time-only masked reconstruction.
To the best of our knowledge, only NPC and Mockingjay examine this kind of behavior.
Note that we pre-train Mockingjay and TERA with the same implementation\footnotemark[\value{footnote}] and setting.
The only difference is by adding the frequency masking.

\subsubsection{Learning with Different Acoustic Features}
\label{ssec:exp_analysis_feature}
In Figure~\ref{fig:analysis}, we pre-trained TERA with MFCC and FBANK as both input and output targets, instead of log Mel.
The setting of these acoustic features is identical to the ones listed in Table~\ref{table:upstreams}.
For phoneme classification on both LibriSpeech and TIMIT, pre-training on log Mel outperforms the other two, while pre-training on FBANK is better than MFCC.
By using a more primitive feature, the model can preserve more phoneme information.
However, the results are opposite for keyword spotting, where models pre-trained on MFCC slightly outperforms FBANK, and models pre-trained on FBANK slightly exceeds log Mel.
Learning from all features successfully preserves the speaker information, with the model pre-trained on log Mel achieving the highest score for speaker recognition.
In general, pre-training with log Mel features results in the best performing model.
This aligns with previous work~\cite{apc1, apc2, improved_apc, vq_apc, npc, mockingjay, audioalbert} (See Table~\ref{table:upstreams}) that also uses 80-dim log Mel.

\subsubsection{Effect of Different Network Depth}
\label{ssec:exp_analysis_network_depth}
We present classification accuracy of three TERA models, including base (3-layer), medium (6-layer), large (12-layer) in Figure~\ref{fig:analysis}.
We observe that performance decay for all tasks as model depth increases.
We conclude that a small model is sufficient to solve the proposed pre-training task, which is reasonable as a lot of work also uses 3-layer models (See Table~\ref{table:upstreams}).
Note that here TERA is used for representation extraction and not fine-tune.
In our later ASR experiments, we find that large models outperform small models during fine-tuning.
The above discovery aligns with our previous finding with wav2vec 2.0, that large models are excellent for fine-tuning but not representation extraction.
On the other hand, small models are adequate for feature extraction than large models.

\subsubsection{Comparing TERA with SpecAugment}
\label{ssec:exp_analysis_spec_augment}
SpecAugment is a regularization technique for improving ASR training~\cite{spec_augment}.
It shares a similar ideology with our time and frequency alteration.
We consider the LD Policy of SpecAugment, which is the best performing policy described in the paper.
Following the notations in SpecAugment~\cite{spec_augment}, the LD Policy has the following parameters: time mask parameter \textit{T=100} (maximum length of the consecutive time mask), frequency mask parameter \textit{F=9} (maximum length of the consecutive frequency mask), number of time masks \textit{mT=2} (amount of consecutive mask blocks in time), and number of frequency masks \textit{mF=2} (amount of consecutive mask blocks in frequency).
Here we first point out some key differences between the proposed method and SpecAugment~\cite{spec_augment}, then we present the experiment results of comparing TERA with SpecAugment.

\textit{The difference in time mask length.}
SpecAugment uses a longer time mask length of up to 100 compared to TERA’s length of 7.
In SpecAugment, the time mask length applied is chosen from a uniform distribution from 0 to the maximum consecutive length (\textit{T}) of 100.
The considerable variation of time mask length introduces some problems during pre-training.
Empirically, we find that pre-training loss of reconstruction from SpecAugment is volatile and not stable.
Also, in our experiments, we find that the large variation of time mask length makes it hard for the pre-trained model to encode phonetic information.

\textit{The difference in number of time masks.}
SpecAugment uses two consecutive mask blocks (\textit{mT=2}) for each input. 
A fixed amount of consecutive mask blocks is employed.
Whereas TERA determines the number of mask blocks by the maximum time alteration percentage $P_T=15\%$.
The amount of mask blocks $T_{num}$ is then given as $T_{num} = \nint{ P_T \times L_x \div W_T}$, where $L_x$ is the length of input.
Simply put, the number of masking blocks will change according to different input lengths.
The number of time masks will vastly affect the model's training for long utterances.
By fixing the number of consecutive mask blocks, SpecAugment does not utilize the advantage of longer training samples.

\textit{The difference in masking policy.} 
The SpecAugment masks selected time blocks to zero.
In TERA, following the idea of BERT~\cite{bert}, a more sophisticated policy is applied.
There are three cases for the selected time blocks, 1) mask to zero, 2) replace with random time blocks, and 3) do nothing.
Case 2) helps TERA learn the order of speech and not always reconstruct it from zero, case 3) helps TERA ease the training and testing inconsistency problem.
Overall, TERA uses a more advanced time alteration policy, and SpecAugment uses a simple mask-to-zero method.

\textit{Comparing experimentally.} 
To compare with SpecAugment experimentally, we apply the SpecAugment LD Policy to self-supervised speech training.
We use the same pre-training settings (960 hours) and network architecture for both TERA and SpecAugment (3-layer Transformer Encoders $T_{enc}$ and the 2-layer prediction network $P_{net}$), with only the difference between masking policies discussed above.
We show the results in dark blue and blue dotted color in Figure~\ref{fig:analysis}.
TERA \textit{base 960hr} largely outperforms \textit{SpecAug} on phoneme classification for both LibriSpeech and TIMIT, and on the keyword spotting task.
Also, TERA wins over SpecAugment on the speaker recognition task.
We conclude that SpecAugment is suitable for regularizing ASR training but not self-supervised learning.
On the other hand, TERA is more effective for self-supervised representation learning.

\begin{table}[t]   
    \centering
    \begin{tabular}{|l||c|c|c|c|}
    \hline
    Models & Pre-train & Labels & WER & Rescore\\ \hline \hline
    liGRU + MFCC & None & 100 hr & 8.66 & 6.42 \\ 
    liGRU + FBANK & None & 100 hr & 8.64 & 6.34 \\ 
    liGRU + fMLLR & None & 100 hr & 8.63 & 6.25 \\ \hline \hline
    Bidir-CPC \cite{bidir_cpc} & 960 hr & 96 hr& 14.96 & 9.41 \\ 
    Bidir-CPC \cite{bidir_cpc} & 8000 hr & 96 hr & 13.69 & 8.70\\ \hline
    vq-wav2vec \cite{vq_wav2vec} & 960 hr & 960 hr & 6.2 & - \\ \hline
    wav2vec-large \cite{decoar} & 960 hr & 100 hr & 6.92 & - \\ 
    DeCoAR \cite{decoar} & 960 hr & 100 hr & 6.10 & - \\ \hline \hline
    liGRU + TERA-base & 960 hr & 100 hr & \bf 8.31 & \bf 6.01 \\ 
    liGRU + TERA-medium & 960 hr & 100 hr & 8.37 & 6.05 \\ 
    liGRU + TERA-large & 960 hr & 100 hr & 8.35 & \bf 6.01 \\ \hline
    \end{tabular}
    \caption{\small\textbf{Comparison of recent speech representation approaches for ASR.} All results are from training an ASR system on top of frozen representations, without fine-tuning the pre-trained model. We report WER on the LibriSpeech~\cite{librispeech} \textit{test-clean} subset. \vspace{-5mm}} 
    \label{table:libri_asr_compare}
\end{table}
\vspace{5mm}

\subsection{Speech Representations for ASR}
\label{ssec:exp_asr_representation}
We further apply TERA to speech recognition tasks.
In Table~\ref{table:libri_asr_compare}, we list results of TERA and recent work in terms of WER, where all ASR models are trained on top of frozen representations without fine-tuning.
All TERA models use a combination of \textit{"time+freq+mag"} alteration as the auxiliary objective.
All of the works report WER and LM rescored WER (denoted as Rescore) on the \textit{test-clean} subset of LibriSpeech~\cite{librispeech}.
All methods are pre-trained with 960 hours of LibriSpeech, except for one experiment setup in \cite{bidir_cpc} where it uses 8000 hours of data for pre-training.
All of the work use only the \textit{train-clean-100} for downstream adaption, except for the setup in \cite{bidir_cpc} and \cite{vq_wav2vec} that use 96 hours and 960 hours of label, respectively.
We use the decoding and rescoring setup described in Section~\ref{ssec:asr_setup} for all \textit{liGRU + TERA} and its variation, as well as \textit{liGRU} with baseline features.
Beam-search decoding with a 4-gram language model is used for Bidir-CPC~\cite{bidir_cpc}.
For wav2vec-large and DeCoAR~\cite{decoar}, a 4-gram LM is used in the first-pass decoding.

First, we observe that the model sizes of TERA (i.e., \textit{base}, \textit{medium} and \textit{large}) have little influence on the ASR performance when TERA is used as an extractor for speech representation.
This observation aligns with our previous discovery on other downstream tasks.
The representations from the \textit{base} model are sufficient to improve supervised ASR.
Since all the cited models use different LM setups, it is hard to conclude the WER comparison in Table~\ref{table:libri_asr_compare}.
However, we cite other work's performance to show that the WER achieved in this work is well within the expected range.
We also investigate three baseline features of MFCC, FBANK, and fMLLR.
We use an identical ASR framework and setting of TERA representations for the three features.
Our results suggest that TERA yields constant improvement over surface features in the same ASR framework.

\begin{table}[t]   
    \centering
    \begin{tabular}{|l||c||c|c|c|c|}
    \hline
    Models & Labels & WER & Rescore\\ \hline \hline
    wav2vec 2.0 - large \cite{wav2vec2} & 100 hr & \textbf{2.3} & - \\ \hline
    Discrete BERT + vq-wav2vec \cite{vq_wav2vec_ft} & 100 hr & \bf 4.5 & - \\ 
    Continuous BERT + wav2vec \cite{vq_wav2vec_ft} & 100 hr & 11.8 & - \\ \hline
    Masked Pre-trained Encoders \cite{mpe} & 100 hr & 9.68 & - \\ 
    Masked Pre-trained Encoders \cite{mpe} & \textit{360 hr} & \bf 7.83 & - \\ \hline \hline
    liGRU + TERA-base (fine-tune) & 100 hr & 8.23 & 5.84 \\ 
    liGRU + TERA-medium (fine-tune) & 100 hr & 8.22 & 5.90 \\ 
    liGRU + TERA-large (fine-tune) & 100 hr & \bf 8.00 & \bf 5.80\\ \hline
    MLP + TERA-base (fine-tune) & 100 hr & 8.47 & 6.24 \\ 
    MLP + TERA-medium (fine-tune)& 100 hr & 8.02 & 5.86 \\ 
    MLP + TERA-large (fine-tune)& 100 hr & \bf 7.96 & \bf 5.84 \\ \hline
    \end{tabular}
    \caption{\small\textbf{Comparison of recent pre-training approaches for ASR.} All results are from fine-tuning the pre-trained model as speech encoders as part of the ASR system. ASR WER and WER after LM rescoring on the LibriSpeech~\cite{librispeech} \textit{test-clean} subset are reported. \vspace{-5mm} }
    \label{table:libri_asr_ft_compare}
\end{table}

\subsection{Speech Pre-training for ASR Comparison}
\label{ssec:exp_asr_finetune}
In this section, we compare the results of fine-tuning various pre-trained models for ASR.
All TERA models use a combination of \textit{"time+freq+mag"} alteration as the auxiliary objective.
We summarize the results from previous literature as well as fine-tuning TERA with \textit{liGRU} or \textit{MLP} framework in Table~\ref{table:libri_asr_ft_compare}.
We also list results from recent literature, where all results are from fine-tuning the pre-trained model as an ASR encoder.
Similar to the previous section, we report WER and LM rescored WER on the \textit{test-clean} subset of LibriSpeech~\cite{librispeech}.
The first-pass decoding and LM rescoring setting are described in Section~\ref{ssec:asr_setup}.
All the methods investigated here were pre-trained with 960 hours and use 100 hours of labels, except for Masked Pre-trained Encoders~\cite{mpe} trained with 360 hours of labels.
The wav2vec 2.0~\cite{wav2vec2} uses a Transformer~\cite{transformer} language model with beam search size of 500 for decoding.
The vq-wav2vec~\cite{vq_wav2vec_ft} uses a 4-gram LM during first-pass decoding, and Masked Pre-trained Encoders~\cite{mpe} adopt beam search and RNN LM with CTC decoding.
When fine-tuning TERA with \textit{liGRU} models, performance roughly correlates with the depth of TERA, and the \textit{large} TERA achieved the best WER.
Remember that large models (wav2vec 2.0, TERA-large, etc) did not perform well for feature extraction. 
However, they are effective for ASR fine-tuning.
By comparing the \textit{liGRU} results in Table~\ref{table:libri_asr_compare} and Table~\ref{table:libri_asr_ft_compare}, we see that fine-tuning TERA consistently outperforms the case when TERA is simply used for extraction of speech representation.
The ASR model adopting \textit{base} TERA improves from 6.01\% to 5.84\%, the \textit{medium} TERA improves from 6.05\% to 5.90\%, and the \textit{large} TERA from 6.01\% to 5.80\%.

Here we also cite other work's performance to show that the WER achieved for the proposed method is within the expected range.
The wav2vec 2.0~\cite{wav2vec2} large model achieves a high score of 2.3\%.
However, we argue that it is consists of seven convolution blocks plus 24 transformer blocks.
In contrast, our \textit{base} model contains only a 3-layer Transformer Encoder layers, which brings substantial low-footprint benefits.
The massive wav2vec 2.0 model needs to be trained on 128 V100 GPUs, where the TERA model can be trained on a single GPU.
The discrete BERT + vq-wav2vec~\cite{vq_wav2vec_ft} achieves a high score of 4.5\%.
However, we argue that the two-step pre-training is computation-intensive during model training.
First, a discrete vocabulary of the data is learned from vq-wav2vec~\cite{vq_wav2vec}, and then in the second step a standard BERT~\cite{bert} is trained on these discrete units.
Also, the discrete BERT + vq-wav2vec~\cite{vq_wav2vec_ft} is built by stacking a standard BERT model~\cite{bert} of 12 Transformer Encoder layers~\cite{transformer} on top of vq-wav2vec~\cite{vq_wav2vec}, which consists of an 8-layer encoder network and a 12-layer aggregator network (or context network, as described in \cite{cpc}).
Our small encoder architecture (3-layer) benefits from less computational cost and can run on edge devices during inference for downstream tasks.

We also fine-tune TERA with \textit{MLP} models, and we find a similar trend but sometimes higher WER compared to TERA with \textit{liGRU}.
Using a deeper model with \textit{MLP} gives performance benefit, and \textit{large} achieves the best WER among the \textit{MLP} models.
The reason is that the simple architecture of \textit{MLP} can benefit from a deeper TERA model.
Comparing \textit{MLP} with \textit{liGRU}, \textit{MLP} achieved superior performance than \textit{liGRU} on the \textit{medium} model, and similar performance for the rest of the model size.
Although in general \textit{MLP} outperforms \textit{liGRU}, however \textit{MLP} has the advantage of a fast training and inference time, thanks to the absence of recurrent units.
Additionally, the parameters of the 1-layer \textit{MLP} is significantly less than the 5-layer \textit{liGRU} models.
To conclude this section for our proposed method, using a deeper model increases ASR performance during fine-tuning.

\begin{table}[t]   
    \centering
    \begin{tabular}{|l||c|c|}
    \hline
    Models & Pre-train & PER \\ \hline \hline
    CNN + TD-filterbanks~\cite{tdfilter} & None & 18.0 \\ 
    CNN + HMM~\cite{cnnphone} & None &  16.5 \\ \hline
    liGRU + MFCC~\cite{ligru} & None & 16.7 \\ 
    liGRU + FBANK~\cite{ligru} & None & 15.8 \\ 
    liGRU + fMLLR~\cite{ligru} & None & 14.9 \\ \hline
    wav2vec~\cite{wav2vec} & 80 hr & 17.6 \\ 
    wav2vec~\cite{wav2vec} & 960 hr & 15.6 \\ 
    wav2vec~\cite{wav2vec} & 960 + WSJ 81 hr & 14.7 \\ \hline
    liGRU + TERA-base & 100 hr & 15.2 \\ 
    liGRU + TERA-base & 360 hr & 14.9 \\ 
    liGRU + TERA-base & 460 hr & 14.9 \\ 
    liGRU + TERA-base & 960 hr & \bf 14.5 \\ \hline
    liGRU + TERA-base (fine-tune) & 960 hr & 15.2 \\ 
    MLP + TERA-base (fine-tune) & 960 hr & 16.6 \\ 
    liGRU + TERA-medium & 960 hr & 14.9 \\ \hline
    \end{tabular}
    \caption{\small\textbf{Comparison of pre-training approaches between recent work and the proposed approach on TIMIT~\cite{timit}.} All the pre-training data are from LibriSpeech~\cite{librispeech}, if not specified otherwise. We report testing results in terms of PER. All of the TERA models use the combined auxiliary objective of "time+freq+mag" alteration. \vspace{-5mm} }
    \label{table:timit_asr_compare}
\end{table}

\subsection{Transferring to TIMIT}
\label{ssec:exp_timit}
We then explore how the mismatch of domains between pre-training and downstream tasks affects performance.
For the exploration, we pre-train TERA with LibriSpeech~\cite{librispeech}, and apply the resulting networks to the supervised TIMIT~\cite{timit} ASR task.
The same Hybrid ASR setting and framework described above for LibriSpeech ASR are used, except that we use a learning rate of $4e^{-4}$ and a batch size of $8$.
Testing results of TERA and another self-supervised learning technique, wav2vec~\cite{wav2vec}, are summarized in Table~\ref{table:timit_asr_compare} in terms of PER.
We also list the results of strong supervised systems~\cite{tdfilter, cnnphone, ligru}. 
All of the TERA models use a combination of \textit{"time+freq+mag"} alteration as the auxiliary objective, and are pre-trained with various amounts of data.
As expected, pre-training on a larger amount of data gives performance benefit, and we achieved the best WER (14.5\%) with 960 hours of pre-training data.
We find that for TIMIT ASR as the downstream task, fine-tuning is not helpful, and extracting speech representations from the last layer provides the best performance.
The reason is likely because there is not enough labeled data in TIMIT, which aligns with our discovery in Section~\ref{ssec:exp_fine_tune} for other downstream tasks.
Also, there is no significant gain when extracting features from a larger model \textit{medium}, which aligns with our previous discussion that smaller models are better for feature extraction.

\section{Conclusion}
We propose a novel self-supervised training scheme called TERA, where the model learns from the reconstruction of altered input.
We pre-train TERA using a large amount of unlabeled data,
and adapt TERA to downstream SLP tasks using a limited amount of labeled data.
We demonstrate strong results in phone classification, keyword spotting, speaker recognition, and speech recognition.
We conduct a complete ablation study and a thorough comparison of recent representation learning and pre-training approaches.
We show that TERA pre-trained on one dataset can be easily transferred to another downstream dataset.
We study how self-supervised models behave on more pre-training data and find that time-only masked reconstruction methods cannot benefit from extensive data.
We also study the choice of acoustic features for pre-training.
We show that it plays a crucial role in reconstruction-based self-supervised learning, as various surface features will lead to significantly different downstream performance.
We investigate networks with different depths and find that small models are more suitable for feature extraction than large models.
On the other hand, large models are more effective for fine-tuning than small models.


%



\section*{Acknowledgment}
The authors are grateful to the National Center for High-performance Computing for computer time and facilities.
They thank Shu-wen Yang for implementing a significant part of the S3PRL toolkit and pre-training the APC, VQ-APC, and NPC models; and Yist Y. Lin for implementing the keyword spotting task.

\ifCLASSOPTIONcaptionsoff
  \newpage
\fi


\bibliographystyle{IEEEtran}
\bibliography{ref}

\begin{thebibliography}{10}
\providecommand{\url}[1]{#1}
\csname url@samestyle\endcsname
\providecommand{\newblock}{\relax}
\providecommand{\bibinfo}[2]{#2}
\providecommand{\BIBentrySTDinterwordspacing}{\spaceskip=0pt\relax}
\providecommand{\BIBentryALTinterwordstretchfactor}{4}
\providecommand{\BIBentryALTinterwordspacing}{\spaceskip=\fontdimen2\font plus
\BIBentryALTinterwordstretchfactor\fontdimen3\font minus
  \fontdimen4\font\relax}
\providecommand{\BIBforeignlanguage}[2]{{%
\expandafter\ifx\csname l@#1\endcsname\relax
\typeout{** WARNING: IEEEtran.bst: No hyphenation pattern has been}%
\typeout{** loaded for the language `#1'. Using the pattern for}%
\typeout{** the default language instead.}%
\else
\language=\csname l@#1\endcsname
\fi
#2}}
\providecommand{\BIBdecl}{\relax}
\BIBdecl

\bibitem{cpc}
\BIBentryALTinterwordspacing
A.~van~den Oord, Y.~Li, and O.~Vinyals, ``Representation learning with
  contrastive predictive coding,'' \emph{CoRR}, vol. abs/1807.03748, 2018.
  [Online]. Available: \url{http://arxiv.org/abs/1807.03748}
\BIBentrySTDinterwordspacing

\bibitem{wav2vec}
S.~Schneider, A.~Baevski, R.~Collobert, and M.~Auli, ``wav2vec: Unsupervised
  pre-training for speech recognition,'' \emph{Interspeech}, 2019.

\bibitem{vq_wav2vec}
\BIBentryALTinterwordspacing
A.~Baevski, S.~Schneider, and M.~Auli, ``vq-wav2vec: Self-supervised learning
  of discrete speech representations,'' in \emph{International Conference on
  Learning Representations}, 2020. [Online]. Available:
  \url{https://openreview.net/forum?id=rylwJxrYDS}
\BIBentrySTDinterwordspacing

\bibitem{vq_wav2vec_ft}
A.~{Baevski} and A.~{Mohamed}, ``Effectiveness of self-supervised pre-training
  for asr,'' in \emph{ICASSP 2020 - 2020 IEEE International Conference on
  Acoustics, Speech and Signal Processing (ICASSP)}, 2020, pp. 7694--7698.

\bibitem{wav2vec2}
\BIBentryALTinterwordspacing
A.~Baevski, Y.~Zhou, A.~Mohamed, and M.~Auli, ``wav2vec 2.0: {A} framework for
  self-supervised learning of speech representations,'' in \emph{NeurIPS 2020,
  December 6-12, 2020, virtual}, H.~Larochelle, M.~Ranzato, R.~Hadsell,
  M.~Balcan, and H.~Lin, Eds., 2020. [Online]. Available:
  \url{https://proceedings.neurips.cc/paper/2020/hash/92d1e1eb1cd6f9fba3227870bb6d7f07-Abstract.html}
\BIBentrySTDinterwordspacing

\bibitem{bidir_cpc}
\BIBentryALTinterwordspacing
K.~Kawakami, L.~Wang, C.~Dyer, P.~Blunsom, and A.~van~den Oord, ``Learning
  robust and multilingual speech representations,'' in \emph{Findings of the
  Association for Computational Linguistics: EMNLP 2020}.\hskip 1em plus 0.5em
  minus 0.4em\relax Online: Association for Computational Linguistics, Nov.
  2020, pp. 1182--1192. [Online]. Available:
  \url{https://www.aclweb.org/anthology/2020.findings-emnlp.106}
\BIBentrySTDinterwordspacing

\bibitem{modified_cpc}
M.~Rivi{\`e}re, A.~Joulin, P.-E. Mazar{\'e}, and E.~Dupoux, ``Unsupervised
  pretraining transfers well across languages,'' in \emph{ICASSP 2020-2020 IEEE
  International Conference on Acoustics, Speech and Signal Processing
  (ICASSP)}.\hskip 1em plus 0.5em minus 0.4em\relax IEEE, 2020, pp. 7414--7418.

\bibitem{apc1}
\BIBentryALTinterwordspacing
Y.-A. Chung, W.-N. Hsu, H.~Tang, and J.~Glass, ``{An Unsupervised
  Autoregressive Model for Speech Representation Learning},'' in \emph{Proc.
  Interspeech 2019}, 2019, pp. 146--150. [Online]. Available:
  \url{http://dx.doi.org/10.21437/Interspeech.2019-1473}
\BIBentrySTDinterwordspacing

\bibitem{apc2}
Y.~{Chung} and J.~{Glass}, ``Generative pre-training for speech with
  autoregressive predictive coding,'' in \emph{ICASSP 2020 - 2020 IEEE
  International Conference on Acoustics, Speech and Signal Processing
  (ICASSP)}, 2020, pp. 3497--3501.

\bibitem{improved_apc}
\BIBentryALTinterwordspacing
Y.-A. Chung and J.~Glass, ``Improved speech representations with multi-target
  autoregressive predictive coding,'' in \emph{Proceedings of the 58th Annual
  Meeting of the Association for Computational Linguistics}.\hskip 1em plus
  0.5em minus 0.4em\relax Online: Association for Computational Linguistics,
  Jul. 2020, pp. 2353--2358. [Online]. Available:
  \url{https://www.aclweb.org/anthology/2020.acl-main.213}
\BIBentrySTDinterwordspacing

\bibitem{vq_apc}
Y.-A. Chung, H.~Tang, and J.~Glass, ``Vector-quantized autoregressive
  predictive coding,'' \emph{Interspeech 2020}, pp. 3760--3764, 2020.

\bibitem{decoar}
S.~{Ling}, Y.~{Liu}, J.~{Salazar}, and K.~{Kirchhoff}, ``Deep contextualized
  acoustic representations for semi-supervised speech recognition,'' in
  \emph{ICASSP 2020 - 2020 IEEE International Conference on Acoustics, Speech
  and Signal Processing (ICASSP)}, 2020, pp. 6429--6433.

\bibitem{audio2vec_1}
\BIBentryALTinterwordspacing
M.~Tagliasacchi, B.~Gfeller, F.~de~Chaumont~Quitry, and D.~Roblek,
  ``Self-supervised audio representation learning for mobile devices,''
  \emph{CoRR}, vol. abs/1905.11796, 2019. [Online]. Available:
  \url{http://arxiv.org/abs/1905.11796}
\BIBentrySTDinterwordspacing

\bibitem{audio2vec_2}
M.~{Tagliasacchi}, B.~{Gfeller}, F.~d.~C.~{Quitry}, and D.~{Roblek},
  ``Pre-training audio representations with self-supervision,'' \emph{IEEE
  Signal Processing Letters}, vol.~27, pp. 600--604, 2020.

\bibitem{wavenet_autoencoder}
J.~Chorowski, R.~J. Weiss, S.~Bengio, and A.~van~den Oord, ``Unsupervised
  speech representation learning using wavenet autoencoders,'' \emph{IEEE/ACM
  Transactions on Audio, Speech, and Language Processing}, vol.~27, no.~12, p.
  2041–2053, Dec 2019.

\bibitem{vc}
A.~T. Liu, P.-c. Hsu, and H.-Y. Lee, ``Unsupervised end-to-end learning of
  discrete linguistic units for voice conversion,'' \emph{Interspeech}, Sep
  2019.

\bibitem{phase_predict}
F.~de~Chaumont~Quitry, M.~Tagliasacchi, and D.~Roblek, ``Learning audio
  representations via phase prediction,'' 2019.

\bibitem{pase}
S.~Pascual, M.~Ravanelli, J.~Serrà, A.~Bonafonte, and Y.~Bengio, ``Learning
  problem-agnostic speech representations from multiple self-supervised
  tasks,'' \emph{Interspeech 2019}, Sep 2019.

\bibitem{convDMM}
\BIBentryALTinterwordspacing
S.~Khurana, A.~Laurent, W.-N. Hsu, J.~Chorowski, A.~Ła{\'n}cucki, R.~Marxer,
  and J.~Glass, ``{A Convolutional Deep Markov Model for Unsupervised Speech
  Representation Learning},'' in \emph{{Interspeech 2020}}, Shanghai, China,
  Oct. 2020. [Online]. Available:
  \url{https://hal.archives-ouvertes.fr/hal-02912029}
\BIBentrySTDinterwordspacing

\bibitem{mockingjay}
A.~T. Liu, S.-w. Yang, P.-H. Chi, P.-c. Hsu, and H.-y. Lee, ``Mockingjay:
  Unsupervised speech representation learning with deep bidirectional
  transformer encoders,'' \emph{ICASSP 2020}, May 2020.

\bibitem{audioalbert}
P.-H. Chi, P.-H. Chung, T.-H. Wu, C.-C. Hsieh, S.-W. Li, and H.~yi~Lee, ``Audio
  albert: A lite bert for self-supervised learning of audio representation,''
  in \emph{SLT 2020}, 2020.

\bibitem{speech_encoder}
W.~Wang, Q.~Tang, and K.~Livescu, ``Unsupervised pre-training of bidirectional
  speech encoders via masked reconstruction,'' \emph{ICASSP 2020}, May 2020.

\bibitem{speechxlnet}
\BIBentryALTinterwordspacing
X.~Song, G.~Wang, Y.~Huang, Z.~Wu, D.~Su, and H.~Meng, ``{Speech-XLNet:
  Unsupervised Acoustic Model Pretraining for Self-Attention Networks},'' in
  \emph{Interspeech 2020}, 2020, pp. 3765--3769. [Online]. Available:
  \url{http://dx.doi.org/10.21437/Interspeech.2020-1511}
\BIBentrySTDinterwordspacing

\bibitem{mpc}
\BIBentryALTinterwordspacing
S.~Li, L.~Li, Q.~Hong, and L.~Liu, ``{Improving Transformer-Based Speech
  Recognition with Unsupervised Pre-Training and Multi-Task Semantic Knowledge
  Learning},'' in \emph{Interspeech 2020}, 2020, pp. 5006--5010. [Online].
  Available: \url{http://dx.doi.org/10.21437/Interspeech.2020-2007}
\BIBentrySTDinterwordspacing

\bibitem{mpc2}
\BIBentryALTinterwordspacing
D.~Jiang, W.~Li, R.~Zhang, M.~Cao, N.~Luo, Y.~Han, W.~Zou, and X.~Li, ``A
  further study of unsupervised pre-training for transformer based speech
  recognition,'' in \emph{Submitted to International Conference on Learning
  Representations}, 2021, under review. [Online]. Available:
  \url{https://openreview.net/forum?id=hrpSB_rzQTU}
\BIBentrySTDinterwordspacing

\bibitem{mpe}
L.~Liu and Y.~Huang, ``Masked pre-trained encoder base on joint
  ctc-transformer,'' 2020.

\bibitem{npc}
A.~H. Liu, Y.-A. Chung, and J.~Glass, ``Non-autoregressive predictive coding
  for learning speech representations from local dependencies,'' 2020.

\bibitem{transformer}
A.~Vaswani, N.~Shazeer, N.~Parmar, J.~Uszkoreit, L.~Jones, A.~N. Gomez,
  u.~Kaiser, and I.~Polosukhin, ``Attention is all you need,'' in
  \emph{Proceedings of the 31st International Conference on Neural Information
  Processing Systems}, ser. NIPS'17.\hskip 1em plus 0.5em minus 0.4em\relax Red
  Hook, NY, USA: Curran Associates Inc., 2017, p. 6000–6010.

\bibitem{nlp_transfer2}
C.~Sun, X.~Qiu, Y.~Xu, and X.~Huang, ``How to fine-tune bert for text
  classification?'' \emph{Chinese Computational Linguistics}, p. 194–206,
  2019.

\bibitem{nlp_transfer3}
A.~Chronopoulou, C.~Baziotis, and A.~Potamianos, ``An embarrassingly simple
  approach for transfer learning from pretrained language models,''
  \emph{Proceedings of the 2019 Conference of the North}, 2019.

\bibitem{S3PRL}
\BIBentryALTinterwordspacing
A.~T. Liu and Y.~Shu-wen, ``The {S3PRL} toolkit: Self-supervised speech
  pre-training and representation learning,'' 2020. [Online]. Available:
  \url{https://github.com/s3prl/s3prl}
\BIBentrySTDinterwordspacing

\bibitem{bert}
\BIBentryALTinterwordspacing
J.~Devlin, M.-W. Chang, K.~Lee, and K.~Toutanova, ``{BERT}: Pre-training of
  deep bidirectional transformers for language understanding,'' in
  \emph{Proceedings of the 2019 Conference of the North {A}merican Chapter of
  the Association for Computational Linguistics: Human Language Technologies,
  Volume 1 (Long and Short Papers)}.\hskip 1em plus 0.5em minus 0.4em\relax
  Minneapolis, Minnesota: Association for Computational Linguistics, Jun. 2019,
  pp. 4171--4186. [Online]. Available:
  \url{https://www.aclweb.org/anthology/N19-1423}
\BIBentrySTDinterwordspacing

\bibitem{roberta}
\BIBentryALTinterwordspacing
Y.~Liu, M.~Ott, N.~Goyal, J.~Du, M.~Joshi, D.~Chen, O.~Levy, M.~Lewis,
  L.~Zettlemoyer, and V.~Stoyanov, ``Roberta: {A} robustly optimized {BERT}
  pretraining approach,'' \emph{CoRR}, vol. abs/1907.11692, 2019. [Online].
  Available: \url{http://arxiv.org/abs/1907.11692}
\BIBentrySTDinterwordspacing

\bibitem{ctc}
A.~Graves, S.~Fern{\'a}ndez, F.~Gomez, and J.~Schmidhuber, ``Connectionist
  temporal classification: labelling unsegmented sequence data with recurrent
  neural networks,'' in \emph{Proceedings of the 23rd international conference
  on Machine learning}, 2006, pp. 369--376.

\bibitem{rnn}
T.~Mikolov, M.~Karafi{\'a}t, L.~Burget, J.~{\v{C}}ernock{\`y}, and
  S.~Khudanpur, ``Recurrent neural network based language model,'' in
  \emph{Eleventh annual conference of the international speech communication
  association}, 2010.

\bibitem{elmo}
M.~Peters, M.~Neumann, M.~Iyyer, M.~Gardner, C.~Clark, K.~Lee, and
  L.~Zettlemoyer, ``Deep contextualized word representations,''
  \emph{Proceedings of the 2018 Conference of the North American Chapter of the
  Association for Computational Linguistics: Human Language Technologies,
  Volume 1 (Long Papers)}, 2018.

\bibitem{albert}
\BIBentryALTinterwordspacing
Z.~Lan, M.~Chen, S.~Goodman, K.~Gimpel, P.~Sharma, and R.~Soricut, ``Albert: A
  lite bert for self-supervised learning of language representations,'' in
  \emph{International Conference on Learning Representations}, 2020. [Online].
  Available: \url{https://openreview.net/forum?id=H1eA7AEtvS}
\BIBentrySTDinterwordspacing

\bibitem{xlnet}
Z.~{Yang}, Z.~{Dai}, Y.~{Yang}, J.~{Carbonell}, R.~{Salakhutdinov}, and Q.~V.
  {Le}, ``{XLNet: Generalized Autoregressive Pretraining for Language
  Understanding},'' \emph{arXiv e-prints}, p. arXiv:1906.08237, Jun. 2019.

\bibitem{mockingjay_defense}
\BIBentryALTinterwordspacing
H.~Wu, A.~T. Liu, and H.~yi~Lee, ``{Defense for Black-Box Attacks on
  Anti-Spoofing Models by Self-Supervised Learning},'' in \emph{Proc.
  Interspeech 2020}, 2020, pp. 3780--3784. [Online]. Available:
  \url{http://dx.doi.org/10.21437/Interspeech.2020-2026}
\BIBentrySTDinterwordspacing

\bibitem{understanding}
\BIBentryALTinterwordspacing
S.~wen Yang, A.~T. Liu, and H.~yi~Lee, ``{Understanding Self-Attention of
  Self-Supervised Audio Transformers},'' in \emph{Proc. Interspeech 2020},
  2020, pp. 3785--3789. [Online]. Available:
  \url{http://dx.doi.org/10.21437/Interspeech.2020-2231}
\BIBentrySTDinterwordspacing

\bibitem{spec_augment}
D.~S. Park, W.~Chan, Y.~Zhang, C.-C. Chiu, B.~Zoph, E.~D. Cubuk, and Q.~V. Le,
  ``Specaugment: A simple data augmentation method for automatic speech
  recognition,'' \emph{Interspeech 2019}, Sep 2019.

\bibitem{slu_bert}
P.~{Wang}, L.~{Wei}, Y.~{Cao}, J.~{Xie}, and Z.~{Nie}, ``Large-scale
  unsupervised pre-training for end-to-end spoken language understanding,'' in
  \emph{ICASSP}, 2020.

\bibitem{bertphone}
S.~Ling, J.~Salazar, Y.~Liu, and K.~Kirchhoff, ``{BERTphone: Phonetically-aware
  Encoder Representations for Utterance-level Speaker and Language
  Recognition},'' in \emph{Odyssey 2020 The Speaker and Language Recognition
  Workshop}, 2020.

\bibitem{word2vec}
\BIBentryALTinterwordspacing
T.~Mikolov, K.~Chen, G.~Corrado, and J.~Dean, ``Efficient estimation of word
  representations in vector space,'' in \emph{1st International Conference on
  Learning Representations, {ICLR} 2013, Scottsdale, Arizona, USA, May 2-4,
  2013, Workshop Track Proceedings}, Y.~Bengio and Y.~LeCun, Eds., 2013.
  [Online]. Available: \url{http://arxiv.org/abs/1301.3781}
\BIBentrySTDinterwordspacing

\bibitem{librispeech}
V.~{Panayotov}, G.~{Chen}, D.~{Povey}, and S.~{Khudanpur}, ``Librispeech: An
  {ASR} corpus based on public domain audio books,'' in \emph{ICASSP}, 2015.

\bibitem{long}
\BIBentryALTinterwordspacing
N.-Q. Pham, T.-S. Nguyen, J.~Niehues, M.~Müller, and A.~Waibel, ``{Very Deep
  Self-Attention Networks for End-to-End Speech Recognition},'' in
  \emph{Interspeech 2019}, 2019, pp. 66--70. [Online]. Available:
  \url{http://dx.doi.org/10.21437/Interspeech.2019-2702}
\BIBentrySTDinterwordspacing

\bibitem{adamW}
\BIBentryALTinterwordspacing
I.~Loshchilov and F.~Hutter, ``Decoupled weight decay regularization,'' in
  \emph{International Conference on Learning Representations}, 2019. [Online].
  Available: \url{https://openreview.net/forum?id=Bkg6RiCqY7}
\BIBentrySTDinterwordspacing

\bibitem{timit}
J.~S. {Garofolo}, L.~F. {Lamel}, W.~M. {Fisher}, J.~G. {Fiscus}, and D.~S.
  {Pallett}, ``{DARPA TIMIT acoustic-phonetic continous speech corpus CD-ROM.
  NIST speech disc 1-1.1},'' NASA STI/Recon Technical Report N, p. 27403, Feb.
  1993.

\bibitem{kaldi}
D.~Povey, A.~Ghoshal, G.~Boulianne, L.~Burget, O.~Glembek, N.~Goel,
  M.~Hannemann, P.~Motlicek, Y.~Qian, P.~Schwarz, J.~Silovsky, G.~Stemmer, and
  K.~Vesely, ``The kaldi speech recognition toolkit,'' in \emph{ASRU}, 2011.

\bibitem{timit_phone}
C.~Lopes and F.~Perdigao, ``Phone recognition on the timit database,''
  \emph{Speech Technologies/Book}, vol.~1, pp. 285--302, 2011.

\bibitem{speech_commands}
\BIBentryALTinterwordspacing
P.~Warden, ``Speech commands: A public dataset for single-word speech
  recognition.'' \emph{Dataset available online}, 2017. [Online]. Available:
  \url{http://download.tensorflow.org/data/speech_commands_v0.01.tar.gz}
\BIBentrySTDinterwordspacing

\bibitem{hmm}
K.-F. Lee and H.-W. Hon, ``Speaker-independent phone recognition using hidden
  markov models,'' \emph{IEEE Transactions on Acoustics, Speech, and Signal
  Processing}, vol.~37, no.~11, pp. 1641--1648, 1989.

\bibitem{pytorchkaldi}
M.~{Ravanelli}, T.~{Parcollet}, and Y.~{Bengio}, ``The pytorch-kaldi speech
  recognition toolkit,'' in \emph{ICASSP 2019 - 2019 IEEE International
  Conference on Acoustics, Speech and Signal Processing (ICASSP)}, 2019, pp.
  6465--6469.

\bibitem{fmllr}
M.~J. Gales, ``Maximum likelihood linear transformations for hmm-based speech
  recognition,'' \emph{Computer speech \& language}, vol.~12, no.~2, pp.
  75--98, 1998.

\bibitem{librilight}
J.~{Kahn}, M.~{Rivière}, W.~{Zheng}, E.~{Kharitonov}, Q.~{Xu}, P.~E.
  {Mazaré}, J.~{Karadayi}, V.~{Liptchinsky}, R.~{Collobert}, C.~{Fuegen},
  T.~{Likhomanenko}, G.~{Synnaeve}, A.~{Joulin}, A.~{Mohamed}, and E.~{Dupoux},
  ``Libri-light: A benchmark for asr with limited or no supervision,'' in
  \emph{ICASSP 2020}, 2020, pp. 7669--7673,
  \url{https://github.com/facebookresearch/libri-light}.

\bibitem{tdfilter}
N.~Zeghidour, N.~Usunier, I.~Kokkinos, T.~Schaiz, G.~Synnaeve, and E.~Dupoux,
  ``Learning filterbanks from raw speech for phone recognition,'' \emph{ICASSP
  2018}, Apr 2018.

\bibitem{cnnphone}
L.~T{\'o}th, ``Phone recognition with hierarchical convolutional deep maxout
  networks,'' \emph{EURASIP Journal on Audio, Speech, and Music Processing},
  vol. 2015, no.~1, pp. 1--13, 2015.

\bibitem{ligru}
M.~Ravanelli, P.~Brakel, M.~Omologo, and Y.~Bengio, ``Light gated recurrent
  units for speech recognition,'' \emph{IEEE Transactions on Emerging Topics in
  Computational Intelligence}, vol.~2, no.~2, p. 92–102, Apr 2018.

\end{thebibliography}

\footnotesize{
\begin{IEEEbiography}[{\includegraphics[width=0.8in,height=9.in,clip,keepaspectratio]{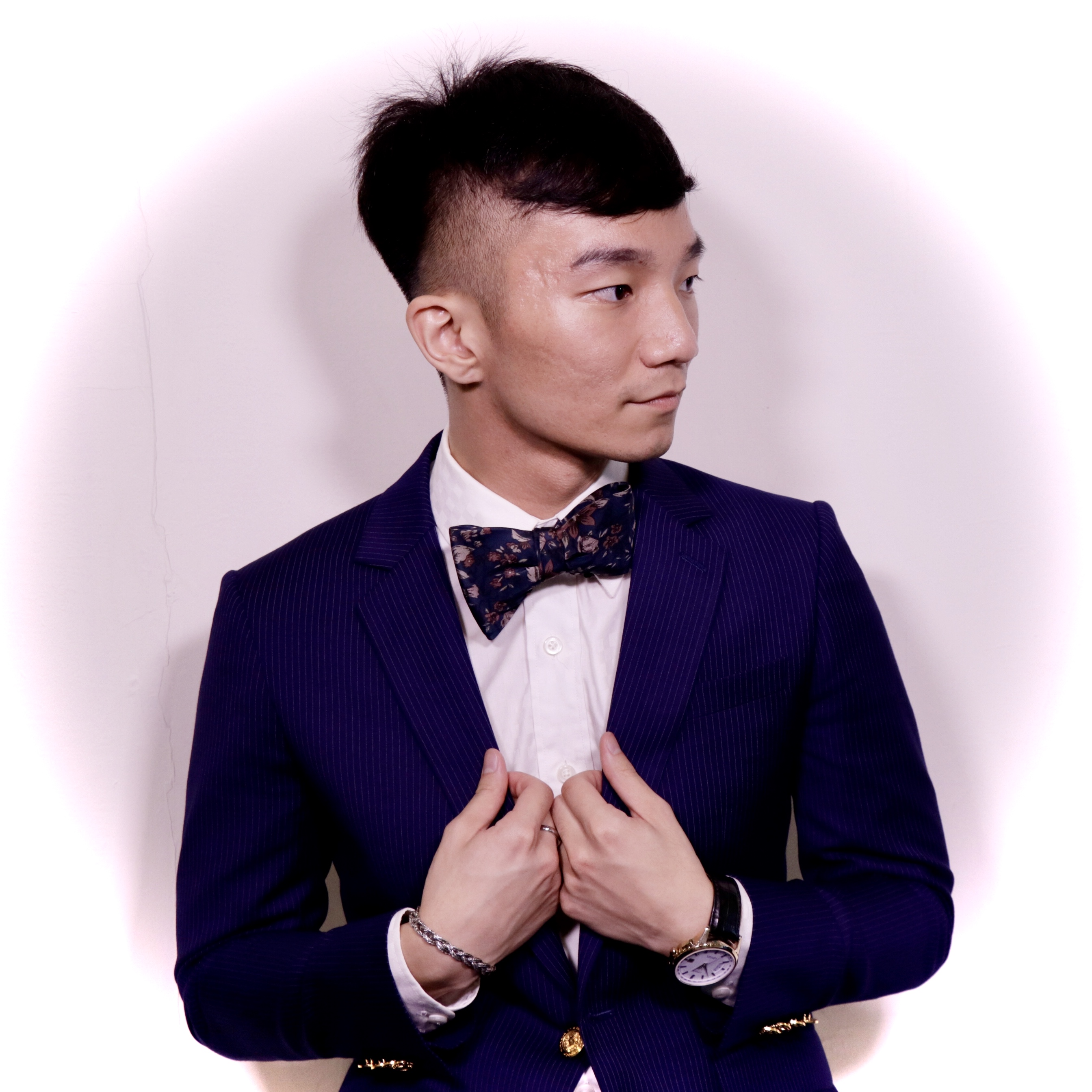}}]{Andy T. Liu} received the bachelor’s degree in electrical engineering from National Taiwan University (NTU), Taipei, Taiwan, in 2018. He is currently working toward the Ph.D. degree with the Graduate Institute of Communication Engineering, NTU, supervised by Professor Hung-yi Lee. His research interests include self-supervised learning, pre-training, and representation learning in the speech and NLP domain.
\end{IEEEbiography}
\vspace{-10mm}

\begin{IEEEbiography}[{\includegraphics[width=1in,height=1.20in,clip,keepaspectratio]{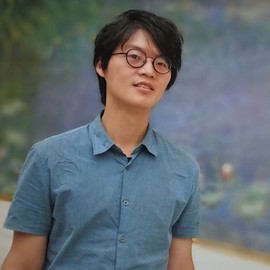}}]{Shang-Wen~Li}
received the Ph.D. degree from MIT Computer Science and Artificial Intelligence Laboratory, Cambridge, MA, USA, in 2016 supervised by Professor Victor Zue. Since 2019, he has been a Senior Applied Scientist with Amazon AWS AI. He was with Apple Siri and Amazon Alexa before joining AWS. His research interests include spoken language understanding, natural language generation, dialog management, and low-resource speech processing.
\end{IEEEbiography}
\vspace{-10mm}

\begin{IEEEbiography}[{\includegraphics[width=1in,height=1.20in,clip,keepaspectratio]{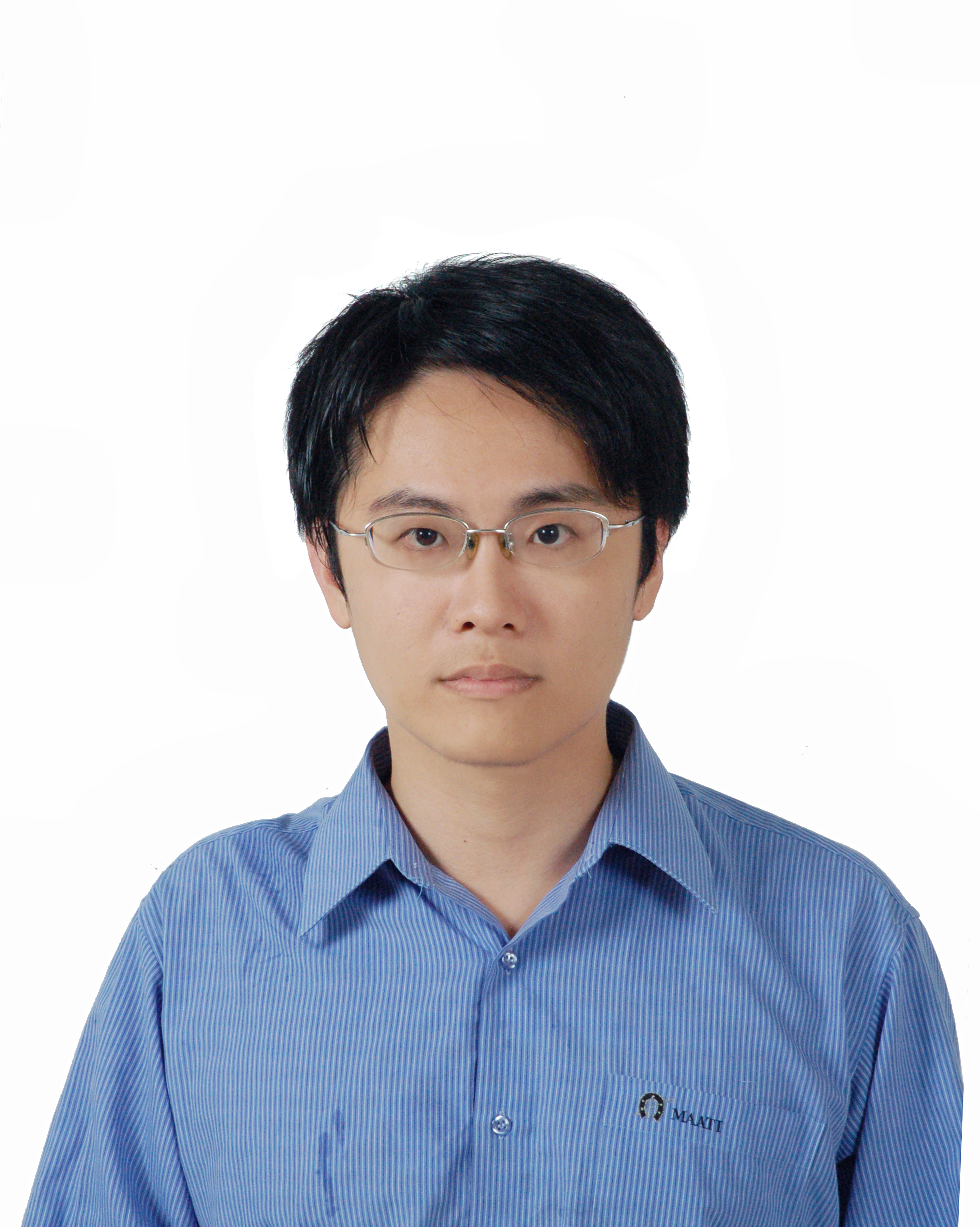}}]{Hung-yi Lee}
received the M.S. and Ph.D. degrees from National Taiwan University, Taipei, Taiwan, in 2010 and 2012, respectively. From September 2012 to August 2013, he was a Postdoctoral Fellow with the Research Center for Information Technology Innovation, Academia Sinica, Taipei, Taiwan. From September 2013 to July 2014, he was a Visiting Scientist with Spoken Language Systems Group, MIT Computer Science and Artificial Intelligence Laboratory, Cambridge, MA, USA. He is currently an Assistant Professor with the Department of Electrical Engineering, National Taiwan University, with a joint appointment with the Department of Computer Science and Information Engineering of the university. His research interests include spoken language understanding, speech recognition, and machine learning.
\end{IEEEbiography}
}
\vfill






\end{document}